\documentclass{article}

\usepackage{microtype}
\usepackage{graphicx}
\usepackage{subcaption}
\usepackage{booktabs} % for professional tables

\usepackage{hyperref}

\usepackage[preprint]{icml2026}

\usepackage{amsmath}
\usepackage{amssymb}
\usepackage{mathtools}
\usepackage{amsthm}

\usepackage{booktabs}
\usepackage{multirow}
\usepackage[table,xcdraw]{xcolor}
\usepackage{color}
\usepackage{makecell}
\usepackage{arydshln}
\usepackage{enumitem}

\usepackage[capitalize,noabbrev]{cleveref}

\theoremstyle{plain}

\theoremstyle{definition}

\theoremstyle{remark}

\usepackage{threeparttable}

\defcitealias{china_gb_45438_2025}{CHN, 2025}
\defcitealias{eu_ai_act_2024}{EU, 2024}
\defcitealias{us_ai_innovation_act_2024}{USA, 2024}

\usepackage[textsize=tiny]{todonotes}

\icmltitlerunning{VocBulwark: Towards Practical Generative Speech Watermarking via Additional-Parameter Injection}

\begin{document}

\twocolumn[
  \icmltitle{VocBulwark: Towards Practical Generative Speech Watermarking \\ via Additional-Parameter Injection}




  \begin{icmlauthorlist}
    % \icmlauthor{Anonymous Author}{aa}
    \icmlauthor{Weizhi Liu}{ecnu}
    \icmlauthor{Yue Li}{hqu}
    \icmlauthor{Zhaoxia Yin}{ecnu}
  %   \icmlauthor{Firstname5 Lastname5}{yyy}
  %   \icmlauthor{Firstname6 Lastname6}{sch,yyy,comp}
  %   \icmlauthor{Firstname7 Lastname7}{comp}
  %   %\icmlauthor{}{sch}
  %   \icmlauthor{Firstname8 Lastname8}{sch}
  %   \icmlauthor{Firstname8 Lastname8}{yyy,comp}
  %   %\icmlauthor{}{sch}
  %   %\icmlauthor{}{sch}
  \end{icmlauthorlist}

  % \icmlaffiliation{aa}{Anonymous information}
  \icmlaffiliation{ecnu}{East China Normal University, Shanghai, China}
  \icmlaffiliation{hqu}{Huaqiao University, Xiamen, China}

  % \icmlcorrespondingauthor{Anonymous Author}{Anonymous Info}
  \icmlcorrespondingauthor{Zhaoxia Yin}{zxyin@cee.ecnu.edu.cn}


  \vskip 0.3in
]



\printAffiliationsAndNotice{}  % no special notice (required even if empty)

\begin{abstract}
  % This document provides a basic paper template and submission guidelines.
  % Abstracts must be a single paragraph, ideally between 4--6 sentences long.
  % Gross violations will trigger corrections at the camera-ready phase.
  Generated speech achieves human-level naturalness but escalates security risks of misuse.
  However, existing watermarking methods fail to reconcile fidelity with robustness, as they rely either on simple superposition in the noise space or on intrusive alterations to model weights.
  To bridge this gap, we propose VocBulwark, an additional-parameter injection framework that freezes generative model parameters to preserve perceptual quality.
  Specifically, we design a Temporal Adapter to deeply entangle watermarks with acoustic attributes, synergizing with a Coarse-to-Fine Gated Extractor to resist advanced attacks.
  Furthermore, we develop an Accuracy-Guided Optimization Curriculum that dynamically orchestrates gradient flow to resolve the optimization conflict between fidelity and robustness.
  Comprehensive experiments demonstrate that VocBulwark achieves high-capacity and high-fidelity watermarking, offering robust defense against complex practical scenarios, with resilience to Codec regenerations and variable-length manipulations.
\end{abstract}

% ==================================================================================================
\section{Introduction}
Speech synthesis has been propelled by Generative Adversarial Networks (GANs)~\cite{goodfellow2014generative} and Diffusion Models~\cite{ho2020denoising, song2020ddim}, achieving fidelity near indistinguishable from human speech. 
However, this realism empowers adversaries to exploit deepfakes for fraud and disinformation, prompting regulators worldwide to enact stringent mandates on traceability~\citepalias {china_gb_45438_2025, us_ai_innovation_act_2024, eu_ai_act_2024}. 
As vocoders serve as the unified waveform generator across heterogeneous pipelines (e.g., Text-to-Speech (TTS)~\cite{xie-etal-2025-towards, ma2025diffusion} and Large Audio Language Models~\cite{cui2025recent, gao2025benchmarking}), watermarking this \textit{pivotal stage} offers a universally effective solution for regulating speech synthesis~\cite{zhao2025sok}.

\begin{figure*}
    \centering
    \includegraphics[width=0.9\textwidth]{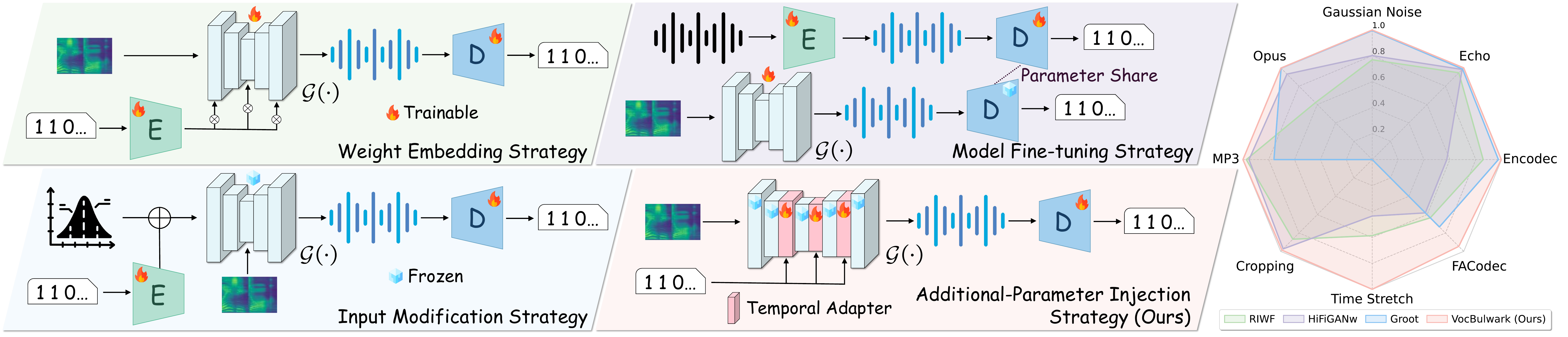}
    \caption{Schematic comparison of generative watermarking paradigms and robustness evaluation.
    The left panel contrasts our \textbf{VocBulwark} (an \textit{Additional-Parameter Injection} strategy) against prevalent baselines, including Weight Embedding, Model Fine-tuning, and Input Modification.
    The right panel visualizes the comparative robustness of these strategies across diverse attack scenarios.}
    \label{fig_overview}
\end{figure*}

Speech watermarking methods have evolved from traditional frequency-domain embedding~\cite{saadi2019normspace, zhao2021desyn-fsvcm, natgunanathan2017pbml} to deep learning-based Encoder-Noise-Decoder (END) frameworks targeting content protection~\cite{chen2023wavmark, san2024audioseal, liu2024timebre, guo2025audiowatermark}.
However, these post-hoc approaches remain limited to content tracking, failing to authenticate the generative source and safeguard model intellectual property (IP).
To bridge this gap, \textit{generative watermarking} has emerged to safeguard model IP by coupling watermarks with the generation process.
As illustrated in Fig.~\ref{fig_overview}, existing watermarking strategies fall into three categories: 
$\mathrm{(i)}$ \textit{Weight Embedding}, modifying internal weights of model~\cite{feng2025riwf},
$\mathrm{(ii)}$ \textit{Model Fine-tuning}, leveraging pre-trained END frameworks to guide fine-tuning~\cite{cheng2024hifi, zhou2024traceablespeech, san2025latent}, 
and $\mathrm{(iii)}$ \textit{Input Modification}, superimposing watermarks onto model inputs~\cite{liu2024groot}.

The aggressive tokenization mechanisms of neural Codecs discretize speech signals while temporal edits invalidate synchronization patterns.
These practical threats precipitate a critical dilemma in current generative watermarking, in which existing paradigms fail to reconcile fidelity and robustness.
Strategies relying on \textit{Input Modification} execute a simple and one-time combination within the noise space, resulting in a superficial addition that is effectively detached from semantic content and thus easily purged by reconstruction processes.
Conversely, \textit{Weight Embedding} and \textit{Model Fine-tuning} fundamentally constitute intrusive alterations to the native parameter space that inevitably disrupt the model's original capability, thereby compromising generation quality to improve watermarking performance.
In contrast, while auxiliary modules thrive in image generation~\cite{feng2024aqualora, lin2024efficient, rezaei2024lawa, ci2025wmadapter, yang2025stableguard}, adapting them to speech is non-trivial due to speech's inherent sensitivity to phase and temporal perturbations.

To overcome these limitations, we propose \textbf{VocBulwark}, an innovative \textit{additional-parameter injection} watermarking method that establishes a reliable regulatory mechanism for heterogeneous generative model architectures, as illustrated in Fig.~\ref{fig_overview}.
We design the \textbf{Temporal Adapter}, a lightweight module that leverages \textit{Acoustic Feature Alignment} and \textit{Frame-level Temporal Broadcasting} to intimately integrate the watermark with intrinsic acoustic attributes while keeping native parameters frozen, thereby ensuring high fidelity while mitigating synchronization dependencies against variable-length attacks.
In coordination with this, our \textbf{Coarse-to-Fine Gated Extractor} utilizes \textit{Multi-scale Feature Aggregation} and \textit{Dual-path Pooling} to capture subtle watermarking artifacts for robust recovery under severe regeneration and distortion.
In addition, addressing the training instability common in multi-objective tasks, we devise an \textbf{Accuracy-Guided Optimization Curriculum} that adaptively adjusts the learning trajectory to harmonize extraction with generation fidelity.
Our contributions can be summarized as follows:
\begin{itemize}
    \item We propose VocBulwark, an innovative additional-parameter injection watermarking framework that establishes a reliable, low-overhead mechanism for model provenance and content regulation across heterogeneous generative architectures.

    \item We propose a Temporal Adapter that uses Acoustic Feature Alignment and Frame-level Broadcasting to achieve temporally invariant embedding in the acoustic space, alongside a Coarse-to-Fine Gated Extractor that leverages Multi-scale Aggregation and Dual-path Pooling to ensure accurate extraction.

    \item We devise an Accuracy-Guided Optimization Curriculum that resolves the multi-objective conflict via a simple-to-complex trajectory, progressively shifting the focus from watermark establishment to fidelity refinement to balance the trade-off between recovery accuracy and perceptual quality.

    \item Comprehensive experiments demonstrate that our method yields high-quality watermarked speech while exhibiting superior robustness against variable-length attacks and neural Codec regeneration attacks.
    
\end{itemize}

% ==================================================================================================
\section{Related Work}

\textbf{Post-hoc Watermarking.}
\label{sec_phw}
Existing approaches can be categorized into handcrafted-based and deep learning-based methods.
The former relied on handcrafted algorithms embedding signals into frequency domains for copyright protection~\cite{saadi2019normspace, zhao2021desyn-fsvcm, natgunanathan2017pbml}.
Conversely, deep learning-based methods have significantly bolstered defense capabilities.
Specifically, \citet{chen2023wavmark} achieved watermark localization via invertible neural networks, while \citet{san2024audioseal} leveraged Codecs for proactive defense against voice cloning.
Furthermore, \citet{liu2024timebre} introduced timbre watermarking for cloning detection, and \citet{liu2023dear} proposed an END framework resilient to re-recording.
While effective against signal-level attacks, these approaches remain confined to \textit{content attribution}, whereas our work extends protection to \textit{model provenance}, establishing a reliable mechanism for safeguarding IP.

\textbf{Generative Watermarking.}
\label{sec_gw}
Regarding Weight Embedding, \citet{feng2025riwf} utilized mask training to identify convolutional kernels prior to the actual watermark embedding.
In the realm of Model Fine-tuning, \citet{zhou2024traceablespeech} pre-trained the watermarked tokenizer of the model to guide the subsequent entire model fine-tuning.
In parallel, \citet{san2025latent} and \citet{cheng2024hifi} leveraged pre-trained END frameworks to watermark training data or guide the fine-tuning process, respectively.
For Input Modification, \citet{liu2024groot} superimposed the encoded watermark latent variable directly onto the model input for synthesis.
In contrast, our VocBulwark fuses the watermark into the acoustic space via TA, guaranteeing intrinsic robustness against distortions.
Crucially, by implementing this via additional-parameter injection under frozen native parameters, our method preserves the perceptual quality.

% ==================================================================================================
\section{Threat Model}
We consider a threat model involving three primary entities: the User, the Model Provider, and the Adversary.

\textbf{Users' Ability and Goal.} 
Users interact with the platform to create speech content.
Specifically, a user queries the TTS service with a text prompt $\mathbf{m}$. The acoustic model $\mathcal{M}$ first converts $\mathbf{m}$ into acoustic features $\mathbf{c} = \mathcal{M}(\mathbf{m})$ (e.g., mel-spectrograms). Subsequently, the vocoder $\mathcal{G}_{\psi}$ synthesizes the raw waveform $\overline{\mathbf{s}} = \mathcal{G}_{\psi}(\mathbf{m}) \in \mathbb{R}^L$, where $L$ denotes the number of samples.
This process can be formulated as:
\begin{equation}
    \overline{\mathbf{s}} = \text{TTS}(\mathbf{m}) = \mathcal{G}_{\psi}(\mathcal{M}(\mathbf{m})) \in \mathbb{R}^L,
\end{equation}
The user's primary objective is to obtain high-fidelity speech $\overline{\mathbf{s}}$ for personal or commercial applications.

\textbf{Model Provider's Ability and Goal.}
To establish model provenance and enforce content regulation, the Model Provider seeks to reliably bind disseminated content to its source model.
The provider assigns a unique watermark identifier $\mathbf{w}$ (e.g., model information or license ID) to each specific model instance utilizing the watermarking algorithm: $\hat{\mathbf{s}} = \mathcal{G}_{\psi}(\mathbf{c}, \mathbf{w})$. 
Consequently, speech generated by the model is intended to carry an identifiable watermark associated with its source.
Even after dissemination, the provider can employ the extraction algorithm $f_{D}: \hat{\mathbf{s}} \rightarrow \hat{\mathbf{w}}$ to recover $\hat{\mathbf{w}}$ from the content $\hat{\mathbf{s}}$.
This mechanism grants the provider the capability to trace the provenance of speech synthesis, thereby facilitating copyright assertion and accountability enforcement against misuse.

\textbf{Adversary's Ability and Goal.}
The adversary aims to evade provenance tracking by removing or obfuscating the watermark $\mathbf{w}$ from the intercepted speech $\hat{\mathbf{s}}$.
We assume a strict \textit{black-box} setting, where the adversary has no access to or knowledge of the model parameters, architecture, or the watermarking algorithm.
They can only manipulate the disseminated speech $\hat{\mathbf{s}}$. 
We define four attack scenarios:
\begin{itemize}
    \item \textbf{Common Attack.}
    The adversary utilizes standard post-processing operations (e.g., Gaussian noise, low-pass filter, and echo) to disrupt the watermark pattern. This is formulated as $\tilde{\mathbf{s}} = \mathcal{A}_{\text{com}}(\hat{\mathbf{s}})$.

    \item \textbf{Variable-length Attack.}
    To disrupt temporal synchronization, the adversary applies time-stretching or cropping, resulting in a variable-length output:
    \begin{equation}
        \tilde{\mathbf{s}} = \mathcal{A}_{\text{VL}}(\hat{\mathbf{s}}) \in \mathbb{R}^{\hat{L}}, \quad \text{where} \quad \hat{L} \ne L.
    \end{equation}

    \item \textbf{Codec Attack.}
    The adversary compresses or regenerates the speech using traditional or neural Codecs, introducing quantization and reconstruction artifacts:
    \begin{equation}
        \tilde{\mathbf{s}} = \mathcal{A}_{\text{Codec}}(\hat{\mathbf{s}}) = \text{Dec}(\text{Enc}(\hat{\mathbf{s}})).
    \end{equation}

    \item \textbf{Compound Attack.}
    The adversary employs a sequential combination of the above operations (e.g., $\tilde{\mathbf{s}} = \mathcal{A}_{\text{Codec}}(\mathcal{A}_{\text{com}}(\hat{\mathbf{s}}))$) to subject the watermark to aggressive, multi-stage distortion.
    
\end{itemize}

\begin{figure*}
    \centering
    \resizebox{0.95\textwidth}{!}{
    \includegraphics{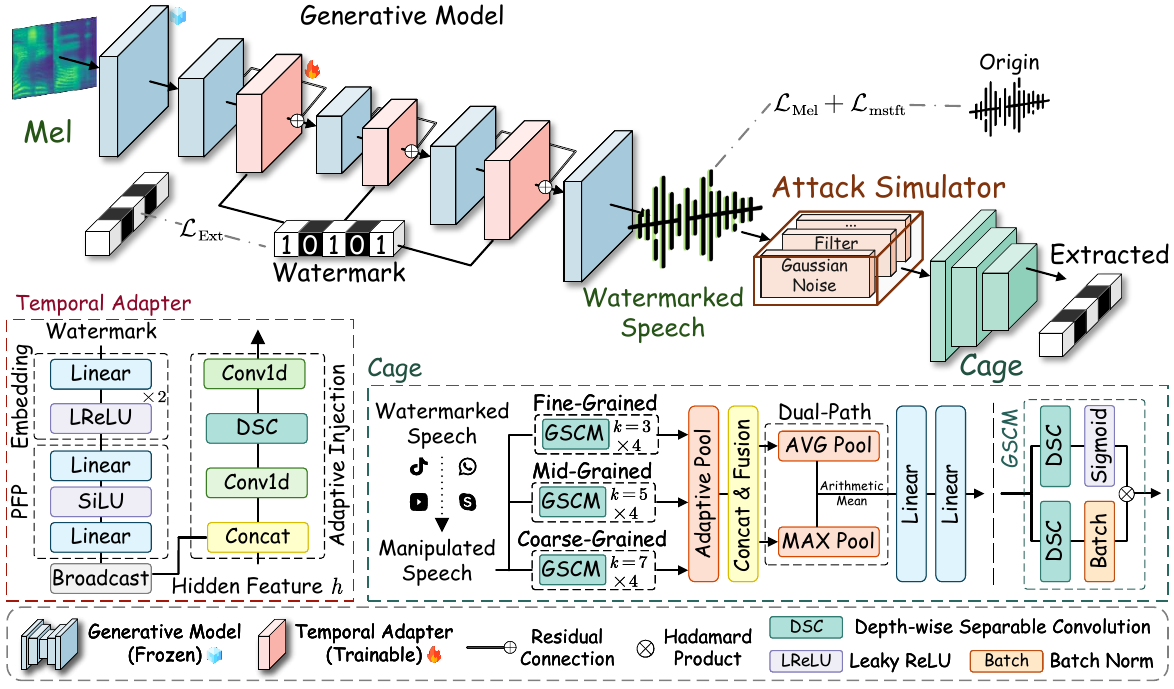}
    }
    \caption{The overall framework of VocBulwark.
    The Temporal Adapter functions as a lightweight module that seamlessly entangles watermarks into acoustic attributes without disrupting the native generation process.
    Following the Attack Simulator, the Cage achieves precise watermark recovery via a coarse-to-fine feature aggregation mechanism.
    The Accuracy-Guided Optimization Curriculum is employed to adaptively modulate the trainable components, optimizing the trade-off between generation fidelity and extraction accuracy.
    }
    \label{fig_pipe}
\end{figure*}

% ==================================================================================================
\section{Methodology}
\subsection{Overview}
The proposed VocBulwark incorporates two synergistic components, as illustrated in Fig.~\ref{fig_pipe}.
On the generation side, the Temporal Adapter (TA) serves as a lightweight injection module. It seamlessly entangles the watermark with acoustic features without disrupting the native workflow of the frozen generative backbone.
On the extraction side, the Coarse-to-Fine Gated Extractor (Cage) adopts the gated separable convolution to precisely recover the watermark from the synthesized speech.
During training, an Attack Simulator is deployed to fortify robustness against real-world distortions. 
Furthermore, we employ an Accuracy-Guided Optimization Curriculum to harmonize conflicting loss terms, thereby facilitating efficient convergence.

\subsection{Watermarking Generative Model}
To internalize the watermark as an intrinsic and time-invariant acoustic representation during the generation process, we design the Temporal Adapter (TA), which incorporates acoustic feature alignment, frame-level temporal broadcasting, and an adaptive injection mechanism, as shown in Fig.~3, with detailed structure provided in App.~\ref{sec_app_ta_archi}

\textbf{Acoustic Feature Alignment.}
The alignment process initiates with an embedding layer $Emb(\cdot)$ encoding the binary watermark $\mathbf{w} \in \{0, 1\}^l$ into a high-dimensional representation.
To mitigate inherent feature redundancy, we construct the Progressive Feature Projection (PFP) $f_\text{Proj}(\cdot)$ to extract intrinsic watermark components.
Specifically, $f_\text{Proj}(\cdot)$ compresses the representation into a compact space and subsequently maps it to match the acoustic dimension $C$ of the hidden features $h \in \mathbb{R}^{B \times C \times T}$ (where $B$ and $T$ denote batch size and temporal length).
This PFP minimizes overhead while distilling intrinsic watermark components.
The resulting projected feature $\mathbf{w}_{\text{proj}}$ is formulated as:
\begin{equation}
\mathbf{w}_{\text{proj}} = f_\text{Proj}(Emb(\mathbf{w})) \in \mathbb{R}^{B \times C \times 1}.
\end{equation}

\textbf{Frame-level Temporal Broadcasting.}
To confer resilience against length-varying manipulations, we employ the broadcasting mechanism $f_\text{BC}(\cdot)$ to replicate $\mathbf{w}_{\text{proj}}$ across the entire temporal axis of $h$, establishing a temporal-invariant injection field whose ubiquity renders the watermark inherently resilient to desynchronization. The broadcasted feature $\mathbf{w}_{\text{latent}}$ can be expressed as:
\begin{equation}
\mathbf{w}_{\text{latent}} = f_\text{BC}(\mathbf{w}_{\text{proj}}) \in \mathbb{R}^{B \times C \times T}.
\end{equation}

\textbf{Adaptive Injection.}
To fuse the watermark within the generation process, we first synthesize a unified feature by concatenating $\mathbf{w}_{\text{latent}}$ with the host features $h$, followed by a downsampling layer $f_{\text{down}}$ to project them into a joint representation.
Afterward, a Depth-wise Separable Convolution (DSC)~\cite{chollet2017dsc} $f_{\text{dsc}}$ processes this feature to facilitate semantic-aware fusion, while a zero convolution $f_{\text{zero}}$ is applied to ensure stable optimization.
Finally, we obtain the watermarked representation $\hat{h}$ via a residual connection, which leverages the identity shortcut to inherently preserve the original semantics and minimize interference:
\begin{equation}
\hat{h} = f_{\text{zero}}(f_{\text{dsc}}(f_{\text{down}}(\mathbf{w}_{\text{latent}} \oplus h))) + h.
\end{equation}

\subsection{Extracting Watermark}
\label{sec_ext}
We propose the Coarse-to-Fine Gated Extractor (Cage) to robustly disentangle the watermark from manipulated speech. Built upon a Gated Separable Convolution Module (GSCM) backbone, Cage integrates a Coarse-to-Fine Feature Aggregation and Dual-path Pooling mechanism, as shown in Fig.~3 (details in App.~\ref{sec_app_cage_archi}).

\textbf{Gated Separable Convolution Module}.
As the core building block of Cage, the gated separable convolution module (GSCM) addresses the challenge of distinguishing subtle watermark features from dominant speech semantics. GSCM $f_{\text{gscm}}(\cdot)$ employs a gating mechanism~\cite{dauphin2017gatedcnn}: a content branch captures local features via DSC, while a parallel gating branch generates a soft mask to adaptively highlight watermark-salient regions while suppressing noise. This backbone endows the extractor with the capability to perform precise, adaptive feature selection within the deep acoustic space.

\textbf{Coarse-to-Fine Feature Aggregation.}
Since attacks induce heterogeneous distortions across varying scales, this mechanism is designed to mitigate localized signal corruption via multi-resolution analysis.
Specifically, we orchestrate fine-, medium-, and coarse-grained branches constructed from cascading GSCM blocks with kernel sizes of $3$, $5$, and $7$, respectively. 
To handle variable-length speech inputs inherent to real-world scenarios, the multi-scale features extracted by these branches are first standardized via Adaptive Average Pooling (AAP).
These aligned representations, denoted as $\mathbf{x}_{\text{fine}}, \mathbf{x}_{\text{mid}}, \text{and } \mathbf{x}_{\text{coarse}}$, are then fused via channel concatenation and refined through two cascaded GSCM blocks $f_{\text{gscm}}(\cdot)$ to synthesize the unified representation $\mathbf{x}_{\text{fuse}}$:
\begin{equation}
\mathbf{x}_{\text{fuse}} = f_{\text{gscm}}(f_{\text{gscm}}(\mathbf{x}_{\text{fine}} \oplus \mathbf{x}_{\text{mid}} \oplus \mathbf{x}_{\text{coarse}})).
\end{equation}

\textbf{Dual-path Pooling.} 
To derive the final representation, we employ a dual-path pooling strategy that synergizes AAP and Adaptive Max Pooling (AMP).
By fusing AMP's discriminative peaks with AAP's global context, the arithmetic mean effectively mitigates feature dilution while buffering against perturbation sensitivity, ensuring a robust and balanced representation.
Ultimately, the decoder $D(\cdot)$ utilizes this refined feature to reconstruct the final watermark $\hat{\mathbf{w}}$:
\begin{equation}
\hat{\mathbf{w}} = D\big( \frac{\text{AAP}(\mathbf{x}_\text{fuse}) + \text{AMP}(\mathbf{x}_\text{fuse})}{2} \big).
\end{equation}

\textbf{Attack Simulator.} 
To enforce robustness against real-world distortions, we design the Attack Simulator (AS) that operates prior to the extraction stage. In each training iteration, the simulator randomly selects and applies an operation of eight operations: Gaussian noise, band-pass/high-pass filtering, suppression, amplitude scaling, echo, dithering, and EnCodec~\cite{defossez2023high}. To prevent convergence failure, all attacks are configured with conservative intensity levels. 
Further settings are elaborated in the App.~\ref{sec_app_attack}.

\subsection{Accuracy-Guided Optimization Curriculum}
\label{sec_train}
To ensure high-fidelity generation and robust extraction, we perform lightweight watermark training optimizing a composite objective of perceptual and extraction losses, governed by the Accuracy-Guided Optimization Curriculum.

To guarantee auditory fidelity, we first employ the Mel-spectrogram loss to align the spectral envelope of the watermarked speech $\hat{\mathbf{s}}$ with the original $\mathbf{s}$:
\begin{equation}
    \mathcal{L}_{\text{Mel}} = ||\log(\phi(\mathbf{s}) + \epsilon)  - \log(\phi(\hat{\mathbf{s}}) + \epsilon)||_1,
\end{equation}
where $\phi(\cdot)$ denotes the Mel-transformation. To further mitigate phase distortion and resolve time-frequency resolution trade-offs, we incorporate the Multi-Scale STFT loss, which computes spectral distances across $m$ analysis scales:
\begin{equation}
\begin{split}
    \mathcal{L}_{\text{mstft}} & = \sum_{i=1}^{m} (\lambda_{\text{mag}} \| |\mathcal{S}_i(\mathbf{s})| - |\mathcal{S}_i(\hat{\mathbf{s}})| \big\|_1 + \lambda_{\text{log}} \\ 
    &\big\| \log_{10}(|\mathcal{S}_i(\mathbf{s})|^2) - \log_{10}(|\mathcal{S}_i(\hat{\mathbf{s}})|^2) \|_1 ),
\end{split}
\end{equation}
where $\mathcal{S}_i(\cdot)$ represents the STFT operation at the $i$-th scale.
Concurrently, we impose a Binary Cross-Entropy loss to supervise watermark recovery, minimizing the bit error rate between the extracted $\hat{\mathbf{w}}$ and the ground truth $\mathbf{w}$:
\begin{equation}
    \mathcal{L}_{\text{Ext}} = - \sum_{i=1}^{l} \left( w_i \log \hat{w}_i + (1-w_i) \log (1-\hat{w}_i) \right).
\end{equation}

The overall objective is formulated as a weighted sum: 
\begin{equation}
\label{eq_loss}
    \mathcal{L} = \lambda_{\text{Mel}} \mathcal{L}_{\text{Mel}} + \lambda_{\text{mstft}} \mathcal{L}_{\text{mstft}} + \lambda_{\text{Ext}} \mathcal{L}_{\text{Ext}}.
\end{equation}
To mitigate the gradient dominance of perceptual losses during the early training phase, we use the curriculum strategy to adaptively modulate the perceptual weights $\lambda_{\text{Mel}}$ and $\lambda_{\text{mstft}}$ based on the real-time extraction accuracy (ACC):
\begin{equation}
\label{eq_agoc}
    \lambda_{\text{Mel}}, \lambda_{\text{mstft}} = 
    \begin{cases}
        0.1, & \text{if } \text{ACC} < \tau_1, \\
        0.2, & \text{if } \tau_1 \le \text{ACC} < \tau_2, \\
        0.5, & \text{if } \text{ACC} \ge \tau_2.
    \end{cases}
\end{equation}
where $\tau_1=0.9$ and $\tau_2=0.95$ are empirically determined thresholds.
This curriculum prioritizes learning the watermarking task first because watermark injection requires establishing a specific pattern, which would be suppressed by premature high-fidelity constraints.
Once extraction stabilizes, we increase the weights to guide the model towards recovering subtle audio details.
For detailed scheduling and algorithmic procedures, please refer to App.~\ref{sec_app_agoc}.

\begin{figure*}[t]
    \centering
    \includegraphics[width=0.9\textwidth]{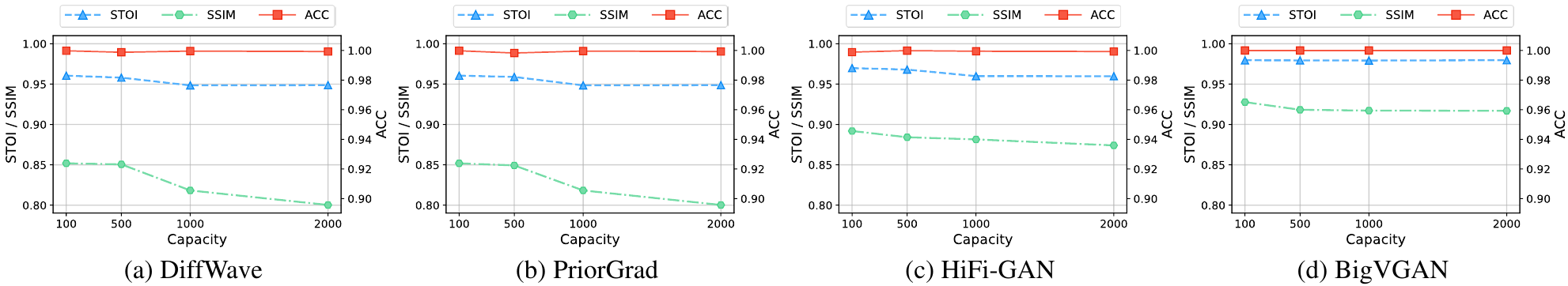}
    \caption{Scalability Analysis Under Varying Watermark Capacities on LJSpeech Dataset.}
    \label{fig_capacity}
\end{figure*}

\begin{table}[t]
\centering
\caption{Comparison of Fidelity on LJSpeech Dataset. The best results are highlighted in \textbf{bold}.}
\resizebox{0.95\linewidth}{!}{
\begin{tabular}{ccccccc}
\toprule
Method (bps)   & STOI~$\uparrow$   & PESQ~$\uparrow$   & SSIM~$\uparrow$   & MCD~$\downarrow$     & DNSMOS~$\uparrow$ & ACC~$\uparrow$    \\ 
\midrule
\rowcolor[HTML]{DDDDDD} 
\multicolumn{7}{c}{Post-hoc Watermarking (Handcrafted-Based)}         \\
Normspace (32) & 0.9689 & 2.8329 & 0.7743 & 21.8547 & 3.1430 & 1.0000 \\
FSVC (32)      & 0.9863 & 3.8095 & 0.9574 & 3.8557  & 3.1778 & 1.0000 \\
PBML (100)     & 0.9984 & 3.9826 & 0.9798 & 0.2063  & 3.2201 & 1.0000 \\
\midrule
\rowcolor[HTML]{DDDDDD} 
\multicolumn{7}{c}{Post-hoc Watermarking (Deep Learning-Based)}       \\
AudioSeal (16) & 0.9971 & 4.3876 & 0.9866 & 0.0090  & 3.2028 & 0.9999 \\
WavMark (32)   & 0.9996 & 4.4794 & 0.9690 & 0.5126  & 3.1316 & 1.0000 \\
TBWM (100)     & 0.9852 & 3.8222 & 0.9388 & 0.7918  & 3.2106 & 1.0000 \\
\midrule
\rowcolor[HTML]{DDDDDD} 
\multicolumn{7}{c}{Generative Watermarking}                           \\
DiffWave       & 0.9655 & 3.5113 & 0.8452 & 6.2890  & 3.1304 & -      \\
RIWF (16)      & 0.9550 & 2.4261 & 0.8160 & \textbf{6.2145}  & 3.0957 & 0.9852 \\
Groot (100)    & 0.9589 & \textbf{3.3871} & 0.8429 & 6.2811  & 3.2726 & 0.9955 \\
\rowcolor[HTML]{f7fcfd} 
Ours[DW] (100)       & \textbf{0.9605} & 3.3398 & \textbf{0.8519} & 6.3682  & \textbf{3.2727} & 0.9998 \\
\cdashline{1-7}
PriorGrad      & 0.9658 & 3.5198 & 0.8454 & 6.2847  & 3.1432 & -      \\
\rowcolor[HTML]{f7fcfd} 
Ours[PG] (100)       & 0.9605 & 3.3398 & 0.8519 & 6.3685  & 3.2727 & 0.9998 \\
\cdashline{1-7}
HiFi-GAN       & 0.9777 & 3.6638 & 0.9434 & 3.7596  & 3.3522 & -      \\
HiFi-GANw (20) & 0.9414 & 2.5862 & \textbf{0.9447} & \textbf{4.8472}  & 2.8718 & 0.9893 \\
\rowcolor[HTML]{f7fcfd} 
Ours[HFG] (100)      & \textbf{0.9698} & \textbf{2.8800} & 0.8919 & 5.7193  & \textbf{3.3266} & 0.9989 \\
\cdashline{1-7}
BigVGAN        & 0.9777 & 3.3397 & 0.9193 & 5.0545  & 3.2044 & -      \\
\rowcolor[HTML]{f7fcfd} 
Ours[BVG] (100)      & 0.9795 & 3.4941 & 0.9277 & 5.0075  & 3.3586 & 1.0000 \\ 
\bottomrule
\end{tabular}
}
\label{tab_fdl}
\end{table}

\begin{table*}[]
\centering
\caption{Comparison of Robustness Against Common and Variable-length Attacks on In-Distribution and Out-of-Distribution Datasets. The best results are highlighted in \textbf{bold} and the second-best are \underline{underlined}.}
\resizebox{0.95\textwidth}{!}{
\begin{tabular}{ccccccccccccccc}
\toprule
 & \multicolumn{10}{c}{Common Attacks}                                                   & \multicolumn{4}{c}{Variable-length Attacks}                            \\ 
\cmidrule(r){2-11}\cmidrule(r){12-15}  
  Method (bps)                & \multicolumn{4}{c}{Gaussian Noise} & LPF    & BPF    & HPF  & Pink Noise   & Echo  & Dither   & \multicolumn{2}{c}{Time Stretch} & \multicolumn{2}{c}{Cropping} \\
\cmidrule(r){2-5}\cmidrule(r){6-8}\cmidrule(r){9-9}\cmidrule(r){10-10}\cmidrule(r){11-11}\cmidrule(r){12-13}\cmidrule(r){14-15}
                              & 5 dB    & 10 dB  & 15 dB  & 20 dB  & 3k     & 0.3-8k & 1k   & 0.2          & Default   & TPDF & 0.9$\times$     & 1.1$\times$    & Front   & Behind   \\
\midrule
\multicolumn{15}{c}{\textit{LJSpeech (In-Distribution Dataset)}}   \\
\midrule
\rowcolor[HTML]{DDDDDD} 
\multicolumn{15}{c}{Post-hoc Watermarking (Handcrafted-Based)}   \\
Normspace (32)                              & 0.5640  & 0.5924 & 0.6121 & 0.6215 & 0.5803 & 0.5308 & 0.4742 & 0.5288     & 0.5708       & 1.0000  & 0.6542          & 0.6896         & 0.4976        & 0.4949 \\
FSVC (32)                                   & 0.6656  & 0.7340 & 0.8116 & 0.8824 & 0.9035 & 0.9873 & 0.8045 & 0.8741     & 0.7872       & 1.0000  & 0.7382          & 0.7217         & 0.5939        & 0.5842 \\
PBML (100)                                  & 0.5576  & 0.6101 & 0.6665 & 0.7170 & 0.6741 & 0.9743 & 0.7506 & 0.7771     & 0.6993       & 0.9821  & 0.5057          & 0.5096         & 0.5067        & 0.5070 \\
\midrule
\rowcolor[HTML]{DDDDDD} 
\multicolumn{15}{c}{Post-hoc Watermarking (Deep Learning-Based)}    \\
AudioSeal (16)                              & 0.5951  & 0.6866 & 0.8082 & 0.9104 & 0.9742 & 0.9998 & 0.8852 & 0.8965     & 0.9629       & 0.9998  & 0.5647          & 0.5063         & 0.9932        & 0.9998 \\
WavMark (32)                                & 0.5056  & 0.5181 & 0.5616 & 0.6513 & 0.9999 & 0.9995 & 1.0000 & 0.9043     & 0.8575       & 1.0000  & -               & -              & -             & -      \\
TBWM (100)                                  & 0.5600  & 0.6333 & 0.7215 & 0.8180 & 0.9938 & 0.9897 & 0.9983 & 0.8820     & 0.9450       & 0.9999  & 0.5020          & 0.4989         & 0.9988        & 0.9986 \\
\midrule
\rowcolor[HTML]{DDDDDD} 
\multicolumn{15}{c}{Generative Watermarking}      \\
RIWF (16)                                   & 0.7690  & 0.8752 & 0.9508 & 0.9791 & 0.9871 & 0.9255 & 0.6984 & 0.5925     & 0.9462       & 0.9858  & 0.5860          & 0.5872         & 0.8675        & 0.6388 \\
HiFi-GANw (20)                              & 0.6176  & 0.8013 & 0.9260 & 0.9702 & 0.9144 & \underline{0.9743} & 0.9827 & 0.9801     & 0.9825       & 0.9893  & 0.4348          & 0.5254         & 0.9729        & 0.9807 \\
Groot (100)                                 & \underline{0.9913}  & \underline{0.9939} & 0.9965 & 0.9937 & 0.9866 & \textbf{0.9939} & 0.7633 & \textbf{0.9961}     & 0.9867       & 0.9956  & -               & -              & -             & -      \\
\rowcolor[HTML]{f7fcfd} 
Ours[DW] (100)             & \textbf{0.9986}         & \textbf{0.9986}        & \textbf{0.9993}   & \underline{0.9993} & \textbf{0.9998} & 0.8317 & \underline{0.9889} & \underline{0.9893}     & \textbf{0.9996}       & 0.9977  & \textbf{0.9994}          & \textbf{0.9992}         & 0.9864        & 0.9839 \\
\rowcolor[HTML]{f7fcfd} 
Ours[PG] (100)             & 0.9172         & 0.9923        & \underline{0.9988}   & \textbf{0.9996} & \textbf{0.9998} & 0.7820 & 0.9731 & 0.9431     & \underline{0.9979}       & \underline{0.9993}  & \underline{0.9977}          & \underline{0.9983}         & 0.9863        & 0.9837 \\
\rowcolor[HTML]{f7fcfd} 
Ours[HFG] (100)                              & 0.9076  & 0.9856 & 0.9916 & 0.9965 & 0.9975 & 0.9029 & 0.8749 & 0.9764     & 0.9950       & 0.9989  & 0.9910          & 0.8621         & \textbf{0.9899}        & \textbf{0.9923} \\
\rowcolor[HTML]{f7fcfd} 
Ours[BVG] (100)                               & 0.8932  & 0.9873 & 0.9968 & 0.9990 & \underline{0.9994} & 0.9463 & \textbf{0.9939} & 0.8915     & 0.9963       & \textbf{0.9998}  & 0.9362          & 0.9479         & \underline{0.9891}        & \underline{0.9903} \\ 
\midrule
\multicolumn{15}{c}{\textit{LibriTTS (OOD Dataset)}}   \\
\midrule
Ours[DW] (100)             & 0.9601         & 0.9764        & 0.9845   & 0.9902 & 0.9814 & 0.8021 & 0.9198 & 0.7905     & 0.9802       & 0.9852  & 0.9829          & 0.9866         & 0.9682        & 0.9698 \\
Ours[PG] (100)            & 0.9601         & 0.9764        & 0.9845   & 0.9902 & 0.9814 & 0.8020 & 0.9197 & 0.7905     & 0.9803       & 0.9850  & 0.9829          & 0.9866         & 0.9683        & 0.9698 \\
Ours[HFG] (100)             & 0.8570         & 0.8722        & 0.9039   & 0.9144 & 0.9165 & 0.8448 & 0.9812 & 0.8644     & 0.9123       & 0.9194  & 0.8605          & 0.8628         & 0.8894        & 0.8985 \\
Ours[BVG] (100)              & 0.9232         & 0.9567        & 0.9622  & 0.9606  & 0.9503 & 0.8955 & 0.7974 & 0.9524     & 0.9486       & 0.9640  & 0.9651          & 0.9365         & 0.9484        & 0.9561 \\ 
\midrule
\multicolumn{15}{c}{\textit{AiShell3 (OOD Dataset)}}   \\
\midrule
Ours[DW] (100)             & 0.9867         & 0.9902        & 0.9935   & 0.9973 & 0.9872 & 0.8616 & 0.9095 & 0.7595     & 0.9938       & 0.9869  & 0.9897          & 0.9830         & 0.9847        & 0.9709 \\
Ours[PG] (100)               & 0.9868         & 0.9901        & 0.9935   & 0.9973 & 0.9872 & 0.8618 & 0.9098 & 0.7594     & 0.9939      & 0.9870  & 0.9897          & 0.9829         & 0.9847        & 0.9709  \\
Ours[HFG] (100)                & 0.8687  & 0.8892 & 0.9372 & 0.9502 & 0.9581 & 0.8601 & 0.8914 & 0.9764     & 0.9446       & 0.9600  & 0.8558          & 0.8362         & 0.9568        & 0.9361 \\
Ours[BVG] (100)                               & 0.8932  & 0.9873 & 0.9968 & 0.9990 & 0.9994 & 0.9463 & 0.9939 & 0.8915     & 0.9963       & 0.9998  & 0.9362          & 0.9479         & 0.9885        & 0.9878 \\ 
\bottomrule
\end{tabular}
}
\label{tab_common_atk}
\end{table*}

\begin{table*}[t]
\centering
\caption{Comparison of Robustness Against Codec Attacks on In-Distribution and Out-of-Distribution Datasets.}
\resizebox{0.95\textwidth}{!}{
\begin{tabular}{ccccccccccccccc}
\toprule
\multirow{3}{*}{} & \multicolumn{6}{c}{Traditional Codec Compression Attack}                     & \multicolumn{8}{c}{Neural Codec Regeneration Attack}                                               \\ 
\cmidrule(r){2-7}\cmidrule(r){8-15}
Method (bps)   & \multicolumn{3}{c}{MP3}     & \multicolumn{3}{c}{Opus}    & \multicolumn{4}{c}{EnCodec}         & HiFiCodec & TiCodec & FACodec  & STCodec \\
\cmidrule(r){2-4}\cmidrule(r){5-7}\cmidrule(r){8-11}\cmidrule(r){12-13}\cmidrule(r){14-14}\cmidrule(r){15-15}
                              & 16 kbps & 32 kbps & 64 kbps & 16 kbps & 32 kbps & 64 kbps & 3 kbps & 6 kbps & 12 kbps & 24 kbps & 4 kbps    & 3 kbps  & 4.8 kbps & 4 kbps  \\
\midrule
\multicolumn{15}{c}{\textit{LJSpeech (In-Distribution Dataset)}}   \\
\midrule
\rowcolor[HTML]{DDDDDD} 
\multicolumn{15}{c}{Post-hoc Watermarking (Handcrafted-Based)}        \\
Normspace (32)                & 0.4971  & 0.4973  & 0.4963  & 0.6394  & 0.6296  & 0.6265  & 0.5830 & 0.6136 & 0.6205  & 0.6247  & 0.5061    & 0.5097  & 0.5268   & 0.5083  \\
FSVC (32)                     & 0.5910  & 0.6075  & 0.6127  & 0.8586  & 0.9701  & 0.9985  & 0.6217 & 0.6603 & 0.7232  & 0.7590  & 0.5599    & 0.5576  & 0.6731   & 0.5842  \\
PBML (100)                    & 0.5813  & 0.6422  & 0.7109  & 0.5774  & 0.6537  & 0.7510  & 0.5086 & 0.5058 & 0.5065  & 0.5069  & 0.5031    & 0.4994  & 0.4966   & 0.4974  \\
\midrule
\rowcolor[HTML]{DDDDDD} 
\multicolumn{15}{c}{Post-hoc Watermarking (Deep Learning-Based)}      \\
AudioSeal (16)                & 0.6042  & 0.5889  & 0.5528  & 0.6717  & 0.8276  & 0.9032  & 0.6872 & 0.6571 & 0.6034  & 0.6069  & 0.5997    & 0.6012  & 0.5843   & 0.5927  \\
WavMark (32)                  & 0.9037  & 0.9963  & 0.9998  & 0.7398  & 0.9790  & 0.9998  & 0.8170 & 0.6962 & 0.4926  & 0.4910  & 0.5001    & 0.5002  & 0.5232   & 0.5037  \\
TBWM (100)                    & 0.8621  & 0.9679  & 0.9917  & 0.7509  & 0.9036  & 0.9914  & 0.8196 & 0.7324 & 0.6288  & 0.6569  & 0.5428    & 0.5247  & 0.5549   & 0.5383  \\
\midrule
\rowcolor[HTML]{DDDDDD} 
\multicolumn{15}{c}{Generative Watermarking}                          \\
RIWF (16)                     & 0.9646  & 0.9727  & 0.9851  & 0.6143  & 0.6206  & 0.6224  & 0.7384 & 0.8031 & 0.8574  & 0.8709  & 0.6799    & 0.6461  & 0.6344   & 0.8534  \\
HiFi-GANw (20)                & 0.8450  & 0.9518  & 0.9869  & 0.8337  & 0.9315  & 0.9469  & 0.5409 & 0.5779 & 0.5762  & 0.5826  & 0.5467    & 0.5650  & 0.5806   & 0.5949  \\
Groot (100)                   & 0.7450  & 0.7592  & 0.7441  & 0.9901  & 0.9958  & 0.9961  & 0.9922 & 0.9904 & 0.9749  & 0.9742  & 0.7324    & 0.7127  & 0.7340   & 0.9609  \\
\rowcolor[HTML]{f7fcfd} 
DiffWave (100)                & \underline{0.9983}  & \underline{0.9988} & \textbf{0.9989} & 0.9945 & 0.9970 & \underline{0.9992} & \textbf{0.9684} & \textbf{0.9954}     & \underline{0.9969}  & \textbf{0.9979}   & \underline{0.9904}    & \underline{0.9174}  & 0.9455  & \underline{0.9521}   \\
\rowcolor[HTML]{f7fcfd} 
PriorGrad (100)               & 0.9979  & \underline{0.9988} & \textbf{0.9989} & 0.9936 & 0.9972 & 0.9988 & \underline{0.9680} & \underline{0.9952}     & \textbf{0.9974}  & \underline{0.9977}   & \textbf{0.9914}    & 0.9171  & \underline{0.9470}  & \textbf{0.9535}   \\
\rowcolor[HTML]{f7fcfd} 
HiFi-GAN (100)                & 0.9076  & 0.9856  & 0.9916  & \underline{0.9965}  & \underline{0.9975}  & 0.9029  & 0.9601 & 0.9937 & 0.9939  & 0.9957  & 0.8518    & 0.8497  & \textbf{0.9793}   & 0.9323  \\
\rowcolor[HTML]{f7fcfd} 
BigVGAN (100)                 & \textbf{0.9984}  & \textbf{0.9903}  & \underline{0.9985}  & \textbf{0.9977}  & \textbf{0.9993}  & \textbf{0.9993}  & 0.9114 & 0.9864 & 0.9924  & 0.9948  & 0.9711    & \textbf{0.9985}  & 0.9345   & 0.9147  \\ 
\midrule
\multicolumn{15}{c}{\textit{LibriTTS (OOD Dataset)}}   \\
\midrule
DiffWave (100)                & 0.9891  & 0.9840 & 0.9862 & 0.9581 & 0.9451 & 0.9496 & 0.9576 & 0.9792     & 0.9796  & 0.9789   & 0.9445    & 0.8570  & 0.8604  & 0.9623   \\
PriorGrad (100)               & 0.9888  & 0.9823 & 0.9853 & 0.9557 & 0.9457 & 0.9485 & 0.9569 & 0.9784     & 0.9815  & 0.9798   & 0.9461    & 0.8606  & 0.8530  & 0.9609   \\
HiFi-GAN (100)                & 0.9173  & 0.9185  & 0.9187  & 0.9053  & 0.9135  & 0.9209  & 0.8186 & 0.9182 & 0.9514  & 0.9569  & 0.8213    & 0.8154  & 0.8626   & 0.7966  \\
BigVGAN (100)                 & 0.8475  & 0.8377  & 0.9294  & 0.9098  & 0.9571  & 0.9651  & 0.8341 & 0.9250 & 0.9506  & 0.9602  & 0.7993    & 0.8129  & 0.9653   & 0.8238  \\ 
\midrule
\multicolumn{15}{c}{\textit{AiShell3 (OOD Dataset)}}   \\
\midrule
DiffWave (100)                & 0.9876  & 0.9901 & 0.9871 & 0.9148 & 0.9206 & 0.9409 & 0.8893 & 0.9516     & 0.9820  & 0.9891   & 0.9836    & 0.9651  & 0.8668  & 0.9612   \\
PriorGrad (100)               & 0.9877  & 0.9879 & 0.9872 & 0.9191 & 0.9186 & 0.9401 & 0.8894 & 0.9513     & 0.9833  & 0.9883   & 0.9839    & 0.9666  & 0.8674  & 0.9623   \\
HiFi-GAN (100)                & 0.9551  & 0.9621 & 0.9604 & 0.9426 & 0.9566 & 0.9592 & 0.8978 & 0.9107     & 0.9016  & 0.9030   & 0.8466    & 0.9148  & 0.7741   & 0.8187  \\
BigVGAN (100)                 & 0.9879  & 0.9943  & 0.9994  & 0.9955  & 0.9991  & 0.9991  & 0.9029 & 0.9721 & 0.9860  & 0.9951  & 0.8066    & 0.7794  & 0.9287   & 0.9337  \\ 
\bottomrule
\end{tabular}
}
\label{tab_Codec_atk}
\end{table*}

% ==================================================================================================
\section{Experiments}

\subsection{Experimental Setting}
\label{sec_setting}
\textbf{Datasets and Models.}
We utilize LJSpeech~\cite{Ito2017ljspeech} for standard training, while extending evaluation to out-of-distribution (OOD) LibriTTS~\cite{zen2019libritts} and AiShell3~\cite{shi2021aishell} to assess generalization.
To verify architectural compatibility, we employ both diffusion-based (DiffWave (DW)~\cite{kong2021diffwave}, PriorGrad (PG)~\cite{lee2022priorgrad}) and GAN-based (HiFi-GAN (HFG)~\cite{kong2020hifi}, BigVGAN (BVG)~\cite{lee2023bigvgan}) vocoders as the base model.
Following previous works~\cite{san2024audioseal, liu2024groot}, speech samples are segmented into 1-second clips at 22.05 kHz.

\textbf{Baseline.}
We benchmark against three \textit{generative watermarking} methods: RIWF~\cite{feng2025riwf}, HiFi-GANw~\cite{cheng2024hifi}, and Groot~\cite{liu2024groot}.
For a comprehensive and fair comparison, we also include \textit{post-hoc} approaches, comprising three handcrafted-based (Normspace~\cite{saadi2019normspace}, FSVC~\cite{zhao2021desyn-fsvcm}, PBML~\cite{natgunanathan2017pbml}) and three \textit{deep learning-based} (AudioSeal~\cite{san2024audioseal}, WavMark~\cite{chen2023wavmark}, TBWM~\cite{liu2024timebre}).

\textbf{Evaluation Metrics.}
We assess watermarking fidelity across three dimensions: 
$(\mathrm{i})$ \textit{Standard Objective Metrics}: Short-Time Objective Intelligibility (STOI)~\cite{taal2010stoi} and Perceptual degradation.
Evaluation of Speech Quality (PESQ)~\cite{recommendation2001pesq} quantifies intelligibility and perceptual degradation.
$(\mathrm{ii})$ \textit{Spectro-temporal Fidelity}: Structural Similarity Index Measure (SSIM)~\cite{wang2004ssim} and Mel-Cepstral Distance (MCD)~\cite{kubichek1993mel} measure structural consistency and spectral envelope distortion. 
$(\mathrm{iii})$ \textit{Neural Naturalness}: DNSMOS~\cite{reddy2022dnsmos} serves as a proxy for human-perceived naturalness.
For extraction performance, Bit-wise Accuracy (ACC) reports the correctly recovered bit rate.

\textbf{Implement Details.}
We optimize the parameters using AdamW with an initial learning rate of $5 \times 10^{-4}$ and weight decay of $1 \times 10^{-4}$.
We employ a Cosine Annealing scheduler that decays the learning rate to $1 \times 10^{-6}$ over 50 epochs, with a batch size of 4.
To ensure training stability, gradient clipping is applied with a maximum norm of 1.0.
All experiments are implemented in PyTorch and trained on a single NVIDIA RTX 4090 GPU with 48GB.
The specific experimental settings are detailed in App.~\ref{sec_app_experiment}.

\subsection{Fidelity and Capacity Analysis}
\label{sec_fi_ca}
\textbf{Fidelity.}
To evaluate the speech quality and accuracy of our method, we benchmark four model variants on the LJSpeech dataset against diverse baselines. 
The results presented in Table~\ref{tab_fdl} demonstrate that watermarking via TA injection guarantees superior generation performance while simultaneously ensuring precise recovery capabilities.
On diffusion models, our DW achieved STOI and SSIM of 0.9605 and 0.8519, respectively, surpassing both RIWF (0.9550 and 0.8160) and Groot (0.9589 and 0.8429). 
Simultaneously, on GAN-based architectures, it attained a PESQ of 2.8800, outperforming HiFi-GANw (2.5862).
These results demonstrate the adaptability of our method across diverse vocoders, confirming its alignment with real-world deployment.

\textbf{Capacity.}
Evaluating scalability across capacities from 100 to 2000 bit per second (bps), we observe that the Exceptional performance illustrated in Fig.~\ref{fig_capacity} is underpinned by our PFP, which distills intrinsic principal components to mitigate feature redundancy.
Benefiting from this compact embedding mechanism, the DW variant sustains robust extraction accuracy ($>99\%$) even at a high payload of 2000 bps, while exhibiting only a marginal SSIM decrease (from $\sim$0.85 to $\sim$0.80) that preserves perceptual intelligibility.
This stable equilibrium extends to other variants (e.g., BVG maintaining STOI $\approx$ 0.97). 
Based on these dual observations, we identify \textit{2000 bps as the optimal upper limit that maximizes capacity} while maintaining a stable balance.

\subsection{Robustness Analysis}
\label{sec_robust}
\textbf{Robustness Against Standard Attacks.}
We first established a rigorous evaluation protocol involving common $\mathcal{A}_{\text{com}}(\cdot)$ and variable-length $\mathcal{A}_{\text{VL}}(\cdot)$ attacks to validate the robustness of VocBulwark and benchmark it against existing methods. 
Specifically, all attacks of this benchmark, including Gaussian and pink noise (GN and PN), low-pass, band-pass, and high-pass filtering (LPF, BPF, and HPF), echo (Ec), dither (Dit), time stretching (TS), and 30\% of cropping, for which the parameter configurations are provided in Table~\ref{tab_common_atk}.
Experimental results reveal that our VocBulwark demonstrates \textit{universally stable robustness} across both attacks.
Specifically, under Gaussian noise in 5 dB, our DW maintains 99.86\% accuracy, marginally surpassing Groot (99.13\%) and outperforming RIWF by over 22\% (76.90\%).
Furthermore, regarding variable-length attacks that typically desynchronize post-hoc watermarks, VocBulwark exhibits exceptional stability, attributed to our \textit{Broadcasting} mechanism.
In particular, our DW achieves 99.94\% ACC under TS (0.9$\times$) and 99.31\% under cropping.
Refer to App.~\ref{sec_app_standard_atk} for more experimental results in large capacity.

\begin{table}[]
\large
\centering
\caption{Comparison of Robustness Against Compound Attacks.}
\resizebox{0.95\linewidth}{!}{
\begin{tabular}{ccccccccc}
\toprule
Method (bps)    & GN+Ec       & LPF+Dit  & PN+GN      & Dit+GN    & Ec+Op     & GN+Enc       & Op+MP3       & Op+Enc        \\
% \cmidrule{2-9}
%                 & 15 dB+Default & 3 kHz+TPDF & 0.2+15 dB & TPDF+15 dB & Default+32 kbps & 15 dB + 12 kbps & 32 kbps+32 kbps & 32 kbps+12 kbps \\
\midrule
\multicolumn{9}{c}{\textit{LJSpeech (In-Distribution Dataset)}}   \\
\midrule
\rowcolor[HTML]{DDDDDD} 
\multicolumn{9}{c}{Post-hoc Watermarking (Handcrafted-Based)}                          \\
Normspace (32)  & 0.4970        & 0.5804     & 0.5754      & 0.5664        & 0.5061    & 0.6016        & 0.4982          & 0.4978          \\
FSVC (32)       & 0.7700        & 0.9018     & 0.7605      & 0.7718        & 0.8274    & 0.6699        & 0.5946          & 0.5610          \\
PBML (100)      & 0.6329        & 0.6663     & 0.6408      & 0.5940        & 0.6254    & 0.5128        & 0.6085          & 0.5046          \\
\midrule
\rowcolor[HTML]{DDDDDD} 
\multicolumn{9}{c}{Post-hoc Watermarking (Deep Learning-Based)}                          \\
AudioSeal (16) & 0.7495        & 0.9743     & 0.5760      & 0.7275        & 0.7376    & 0.9342        & 0.5116          & 0.4504          \\
WavMark (32)    & 0.5059        & 0.9999     & 0.5372      & 0.5443        & 0.7128    & 0.4975        & 0.9345          & 0.4858          \\
TBWM (100)      & 0.6682        & 0.9933     & 0.6610      & 0.6830        & 0.7541    & 0.5686        & 0.8660          & 0.6270          \\
\midrule
\rowcolor[HTML]{DDDDDD} 
\multicolumn{9}{c}{Generative Watermarking}                          \\
RIWF (16)       & 0.7706        & 0.9862     & 0.5760      & 0.8905        & 0.8323    & 0.8428        & 0.6187          & 0.6070          \\
HiFi-GANw (20)  & 0.9247        & 0.9146     & 0.8689      & 0.8289        & 0.9508    & 0.4778        & 0.9208          & 0.5235          \\
Groot (100)     & \underline{0.9854}        & 0.9874     & \textbf{0.9917}      & 0.9878        & 0.9898    & 0.9809        & 0.7623          & 0.7520          \\
\rowcolor[HTML]{f7fcfd} 
DiffWave (100)  & \textbf{0.9960}        & 0.9974     & 0.8656      & \textbf{0.9977}        & \textbf{0.9934}    & \textbf{0.9973}        & 0.9900          & 0.9648          \\
\rowcolor[HTML]{f7fcfd} 
PriorGrad (100) & \textbf{0.9960}        & 0.9974     & 0.8657      & \textbf{0.9977}        & \underline{0.9921}    & \underline{0.9964}        & 0.9917          & 0.9605          \\
\rowcolor[HTML]{f7fcfd} 
HiFi-GAN (100)  & 0.9759        & \textbf{0.9989}     & \underline{0.9471}      & 0.9866        & 0.9893    & 0.9725        & \underline{0.9934}          & \underline{0.9827}          \\
\rowcolor[HTML]{f7fcfd} 
BigVGAN (100)   & 0.9495        & \underline{0.9987}     & 0.9029      & \underline{0.9965}        & 0.9567    & 0.9851        & \textbf{0.9958}          & \textbf{0.9899}          \\ 
\midrule
\multicolumn{9}{c}{\textit{LibriTTS (OOD Dataset)}}   \\
\midrule
DiffWave (100)  & 0.9702        & 0.9683     & 0.8094      & 0.9699        & 0.9690    & 0.9643        & 0.9347          & 0.9015          \\
PriorGrad (100) & 0.9842        & 0.9810     & 0.8405      & 0.9815        & 0.9808    & 0.9781        & 0.9451          & 0.9229          \\
HiFi-GAN (100)  & 0.8936        & 0.9165     & 0.8024      & 0.8918        & 0.9043    & 0.8578        & 0.9102          & 0.8904          \\
BigVGAN (100)   & 0.9161        & 0.9503     & 0.9392      & 0.9545        & 0.8339    & 0.9027        & 0.8923          & 0.9369          \\ 
\midrule
\multicolumn{9}{c}{\textit{AiShell3 (OOD Dataset)}}   \\
\midrule
DiffWave (100)  & 0.9933        & 0.9825     & 0.7704      & 0.9917        & 0.9911    & 0.9827        & 0.9189          & 0.8312          \\
PriorGrad (100) & 0.9933        & 0.9826     & 0.7704      & 0.9917        & 0.9918    & 0.9823        & 0.9192          & 0.8320          \\
HiFi-GAN (100)  & 0.9070        & 0.9455     & 0.7065      & 0.9254        & 0.9328    & 0.9097        & 0.9444          & 0.8768          \\
BigVGAN (100)   & 0.9756        & 0.9991     & 0.9782      & 0.9967        & 0.9840    & 0.9862        & 0.9958          & 0.8869          \\ 
\bottomrule
\end{tabular}
}
\label{tab_compound_atk}
\end{table}

\begin{table}[]
\centering
\large
\caption{Comparison of Computational Complexity.}
\resizebox{0.95\linewidth}{!}{
\begin{tabular}{ccccccc}
\toprule
\multirow{3}{*}{Method (bps)} & \multicolumn{2}{c}{Training Params.~$\downarrow$}      & \multirow{2}{*}{Model Size~$\downarrow$} & \multirow{2}{*}{FLOPs (G)~$\downarrow$}  & \multirow{2}{*}{Infer. Time (ms) $\downarrow$} \\
\cmidrule(r){2-3}
 & Stage 1  & Stage 2 &                             &    &                          \\ 
\midrule
RIWF (16)                     & 0.03 M        & 4.43M         & 67.21 MB                    & 3.52 $ \times 10^6$  & 120.47 $ _{\pm 17.74}$ \\
Groot (100)                   & 250.28 M      & -               & 984.99 MB                   & 3.49 $ \times 10^6$  & 153.32 $ _{\pm 17.37}$ \\
\rowcolor[HTML]{f7fcfd} 
Ours[DW] (100)            & \textbf{1.66 M}        & -               & \textbf{40.06 MB}                     & 3.57 $ \times 10^6$  & 136.49 $ _{\pm 10.68}$ \\
\rowcolor[HTML]{f7fcfd} 
Ours[PG] (100)            & \textbf{1.66 M}        & -               & \textbf{40.06 MB}                     & 3.57 $ \times 10^6$  & 100.77 $ _{\pm 13.40}$ \\
\midrule
HiFi-GANw (20)                & 0.77 M*       & 13.94 M       & 138.73 MB                   & 8.75 $ \times 10^5$  & 15.47 $ _{\pm 8.14}$ \\
\rowcolor[HTML]{f7fcfd} 
Ours[HFG] (100)           & 1.76 M        & -               & 59.99 MB                     & \textbf{6.18}$ \mathbf{\times 10^5} $  & \textbf{13.48}$ \mathbf{_{\pm 2.81}} $ \\
\rowcolor[HTML]{f7fcfd} 
Ours[BVG] (100)           & 1.76 M        & -               & 60.21 MB                     & \textbf{6.18}$ \mathbf{\times 10^5} $  & 55.51 $ _{\pm 7.38}$ \\ 
\bottomrule
\multicolumn{6}{l}{ * Denotes the parameters of the pre-trained END watermarking model.} \\
\end{tabular}
}
\label{tab_compute}
\end{table}

\textbf{Robustness Against Codec Attacks.}
To assess robustness against Codec-based removal $\mathcal{A}_{\text{Codec}}(\cdot)$, we establish a benchmark spanning traditional compression (MP3~\cite{brandenburg1999mp3}, Opus~\cite{valin2012opus}) and diverse neural regeneration. 
Guided by the taxonomy in \citet{Guo25discretespeech}, we select general-purpose acoustic tokenizers (EnCodec with Residual Vector Quantization (RVQ), HiFi-Codec~\cite{yang2023hificodec} with Group-Residual Vector Quantization (GRVQ), and TiCodec~\cite{ren2024ticodec} with RVQ and GVQ), semantic distillation models (STCodec~\cite{zhang2024speechtokenizer} with RVQ), and disentanglement-based frameworks (FACodec~\cite{ju2024naturalspeech} with Factorized VQ) to comprehensively quantify the impact of varying tokenization strategies.
For more details, please refer to App.~\ref{sec_app_codec_config}.

Experimental results in Table~\ref{tab_Codec_atk} demonstrate that VocBulwark achieves exceptional resilience against Codec compression and regeneration.
Under traditional Opus compression, our method achieves a superior average accuracy of 99.69\% across tested bitrates, significantly surpassing RIWF (avg. 61.91\%) and HiFi-GANw (avg. 90.40\%) while edging out the state-of-the-art Groot (avg. 99.40\%).
In addition, VocBulwark demonstrates superior resilience against neural Codec regeneration. 
% Under EnCodec compression, our method achieves an impressive average accuracy of 98.97\% across all bitrates.
% Furthermore, against the remaining Codec, 
Particularly against the diverse remaining Codecs other than EnCodec, our method maintains a robust average ACC of 95.14\%, which is significantly higher than HiFi-GANw (57.18\%), RIWF (70.35\%), and Groot (78.50\%).
This universality stems from our \textit{acoustic feature alignment}, which transcends superficial embedding by coupling the watermark with stable acoustic attributes.
We provide more experiments in App.~\ref{sec_app_codec_atk}.

\textbf{Robustness Against Compound Attacks.}
To assess robustness under rigorous real-world scenarios, we further develop compound attacks, as displayed in Table~\ref{tab_compound_atk}.
It is obvious that VocBulwark demonstrates exceptional resilience that achieves near-perfect recovery rates ($>$99\%) in standard compound attacks, 
Most notably, our advantage is substantial in the most severe scenarios involving double compression, where the watermark undergoes consecutive destructive quantization, and existing baselines suffer catastrophic degradation.
Specifically, while the Groot drops significantly to 76.23\% (Op+MP3) and 75.20\% (Op+Enc), VocBulwark maintains high accuracy, achieving 99.00\% and 96.48\% respectively.
This approximate 20\% performance gap firmly establishes that our deeply coupled acoustic features survive even the rigorous stripping effects of cascaded Codec pipelines.
Please refer to App.~\ref{sec_app_cmp_atk} for more details about Robustness.

\textbf{Summary.}
The superior resilience of our method derives from the synergy of additional-parameter injection, acoustic space embedding, and coarse-to-fine extraction.
Crucially, despite training on merely eight mild distortions, our VocBulwark demonstrates exceptional generalization by extrapolating defense to diverse unseen severe attacks, thereby verifying its practical reliability in rigorous scenarios.

\subsection{Efficiency and Generalizability Analysis}
\label{sec_eff}
\textbf{Computational Efficiency.}
We evaluate the computational efficiency of our method with results summarized in Table~\ref{tab_compute}. 
Trainable Params refers to the count of learnable parameters. Model Size includes both the generative models and the training-specific modules. Inference Time represents the total duration for watermark embedding (including speech generation) and extraction.
Attributed to the addition-parameter injection paradigm, VocBulwark achieves 99.3\% parameter reduction and 24$\times$ storage compression compared to Groot, while maintaining superior inference efficiency for real-time generation.

\textbf{Generalizability.}
We further validate VocBulwark's generalization on the OOD LibriTTS and AiShell3 datasets. Table~\ref{tab_common_atk}, Table~\ref{tab_Codec_atk}, and Table~\ref{tab_compound_atk} demonstrate excellent robustness against four distinct categories of attacks.
This universality is primarily attributed to our TA, which embeds watermarks into intrinsic acoustic attributes and achieves deep integration with the generative process. 
Notably, even on the cross-lingual AiShell3, our approach maintains ideal resilience. 
Particularly against Codec attacks, it continues to demonstrate capabilities that outperform baselines.
The detailed analysis is presented in App.~\ref{sec_app_compute} and App.~\ref{sec_app_genera}.

% ==================================================================================================
\section{Conclusion}
To establish a trustworthy ecosystem for model regulation and provenance, we propose VocBulwark, a generative watermarking framework that leverages an innovative additional-parameter injection strategy. 
Specifically, the TA roots the watermark into acoustic attributes without altering the generative backbone, while the Cage utilizes multi-scale aggregation to precisely disentangle watermark features for robust recovery. 
Augmented by the accuracy-guided optimization curriculum, our method achieves a superior equilibrium between perceptual fidelity and extraction robustness. 
Ultimately, VocBulwark stands as a resilient bulwark, providing a vital technical foundation for responsible and regulated AI-generated speech against evolving threats.

\section*{Impact Statement}
The primary objective of this research is to safeguard model intellectual property and enforce content accountability, thereby establishing a robust and verifiable provenance framework for AI-Generated Content.
This paper proposes a practical generative watermarking framework capable of synthesizing watermarked speech that effectively withstands advanced adversarial threats, including neural Codec regeneration and variable-length manipulations.
However, such generative speech watermarking technology holds the potential for misuse, such as unauthorized surveillance or copyright infringement.
Therefore, we strongly advocate for the establishment of strict ethical guidelines for the deployment of such systems to ensure transparency and informed user consent, while aligning with legal frameworks to prevent the abuse of this technology.

\bibliography{main_ref}
\bibliographystyle{icml2026}

%%%%%%%%%%%%%%%%%%%%%%%%%%%%%%%%%%%%%%%%%%%%%%%%%%%%%%%%%%%%%%%%%%%%%%%%%%%%%%%
%%%%%%%%%%%%%%%%%%%%%%%%%%%%%%%%%%%%%%%%%%%%%%%%%%%%%%%%%%%%%%%%%%%%%%%%%%%%%%%
% APPENDIX
%%%%%%%%%%%%%%%%%%%%%%%%%%%%%%%%%%%%%%%%%%%%%%%%%%%%%%%%%%%%%%%%%%%%%%%%%%%%%%%
%%%%%%%%%%%%%%%%%%%%%%%%%%%%%%%%%%%%%%%%%%%%%%%%%%%%%%%%%%%%%%%%%%%%%%%%%%%%%%%
\clearpage
\newpage
\appendix
\onecolumn
\section{Extended Related Work}
\subsection{More Details of Speech Watermarking}
\textbf{Post-hoc Watermarking.}
As detailed in Sec.~\ref{sec_phw}, the prevailing paradigm in speech watermarking research leverages deep learning, typically employing an Encoder-Noise Layer-Decoder (END) architecture to execute both watermark embedding and extraction, with the primary objective of protecting the copyright of speech content.

WavMark~\cite{chen2023wavmark} leverages an Invertible Neural Network and a shift module, in conjunction with synchronization codes, to enable precise watermark localization, protecting the copyright of both real and synthetic audio.
To counteract audio re-recording (AR) threats, DeAR~\cite{liu2023dear} employs a fully convolutional network, integrating specialized AR modeling within an adversarial training framework to establish robust resilience.
AudioSeal~\cite{san2024audioseal} leverages the EnCodec~\cite{defossez2023high} to embed watermarks into discrete tokens and utilize the decoder for synthesis. By incorporating Time-frequency loudness and Masked sample-level detection losses, it effectively enables proactive defense against voice cloning.
TBWM~\cite{liu2024timebre} leverages frequency-domain features and a repetitive embedding strategy to effectively withstand controllable reconstruction-based removal attacks, thereby safeguarding the copyright of voice timbre.
For voice dataset protection, \citet{guo2025audiowatermark} utilizes the style-transfer generative model with random reference patterns to embed specific timbre attribution, reinforced by bi-level adversarial optimization for enhanced performance.

\textbf{Generative Speech Watermarking.}
As discussed in Sec.~\ref{sec_gw}, contemporary generative watermarking approaches are categorized into three primary strategies, including weight embedding, model fine-tuning, and input modification, which are all designed to safeguard the IP of generative models.

As a weight embedding method, RIWF~\cite{feng2025riwf} initially employs Mask Training to identify convolutional kernel parameters suitable for embedding. Subsequently, it conducts Watermarking Training to embed the watermark into these candidate weights, utilizing a normalization step to preserve model performance.
For model fine-tuning approaches, LatentAWM~\cite{san2025latent} utilizes AudioSeal to watermark the training data for model fine-tuning, achieving zero-bit watermarking that enables a detector to confirm the presence of the watermark in the synthesized audio.
Targeting VALL-E~\cite{chen2025valle}, TraceableSpeech~\cite{zhou2024traceablespeech} initially conducts watermark training on its underlying tokenizer (i.e., EnCodec). Subsequently, it performs full-parameter fine-tuning of the VALL-E using this watermarked tokenizer, thereby achieving robust copyright protection.
In addition, Groot~\cite{liu2024groot}, as an input modification method, implements input modification by mapping the watermark to the latent variable aligned with the input dimensionality using the watermark encoder. By performing element-wise addition on the input, the model is steered to synthesize the watermarked speech.

In contrast, the proposed VocBulwark, functioning as an additional-parameter injection method, generates watermarked speech by injecting watermarks into the acoustic space while freezing the vocoder parameters. 
By leveraging deep entanglement with acoustic attributes, VocBulwark enhances speech quality and provides superior robustness against sophisticated attacks. 
Furthermore, it facilitates easier deployment through significantly fewer trainable parameters.

\subsection{Text-to-Speech Synthesis}
TTS pipelines are typically categorized into cascaded systems and end-to-end models. 
In cascaded frameworks, an acoustic model predicts intermediate representations (e.g., Mel-spectrograms) from text, while a vocoder reconstructs raw waveforms from these features. 
Since our work targets the vocoder stage, we focus on its evolution from early autoregressive~\cite{van2016wavenet} and flow-based~\cite{prenger2019waveglow} methods to advanced GANs~\cite{kong2020hifi, lee2023bigvgan} and Diffusion Models~\cite{kong2021diffwave, lee2022priorgrad}.

WaveNet~\cite{van2016wavenet} pioneered sample-by-sample autoregressive modeling, while WaveGlow~\cite{prenger2019waveglow} achieved faster non-autoregressive synthesis via flow-based structures; notably, both paradigms were leveraged by \citet{chen2021distribution} for distribution-preserving generative steganography.
HiFi-GAN~\cite{kong2020hifi}, characterized by Multi-Receptive Field Fusion and dual discriminators, has become a ubiquitous TTS backbone, prompting specific watermarking designs by \citet{cheng2024hifi} and \citet{feng2025riwf}. Its architecture was subsequently enhanced by BigVGAN~\cite{lee2023bigvgan}, which integrates periodic activations to improve out-of-distribution robustness.
The advent of Denoising Diffusion Probabilistic Model~\cite{ho2020denoising} catalyzed efficient vocoders like DiffWave~\cite{kong2021diffwave} (enabling 6-step synthesis) and PriorGrad~\cite{lee2022priorgrad} (utilizing adaptive priors). Recognizing their superior generative capabilities, \citet{liu2024groot} and \citet{feng2025riwf} developed robust generative watermarking methods specifically for these diffusion-based architectures.

Our approach aims to establish a universal watermarking framework tailored for the two dominant vocoder paradigms: Diffusion Models (DMs) and GANs.
Notably, a structural disparity exists wherein DMs synthesize speech from Gaussian noise conditioned on Mel-spectrograms, while GANs utilize Mel-spectrograms as their sole input.
To reconcile these distinct generative workflows, our proposed TA bridges this gap by injecting watermarks within a shared acoustic space. By adaptively aligning with latent features of varying dimensions, the TA ensures seamless integration, thereby demonstrating robust universality across heterogeneous architectures.
By adaptively aligning with the varying feature dimensions within the vocoder, the TA ensures the effective fusion of watermarks with acoustic attributes, thereby demonstrating robust universality across heterogeneous architectures.

\subsection{Generative Image Watermarking}
In this section, we primarily discuss existing generative image watermarking methods analogous to additional-parameter injection.
LaWa~\cite{rezaei2024lawa} implements a multi-scale embedding strategy preceding the upsampling blocks of the Latent Diffusion Model (LDM) Decoder, wherein the one-dimensional watermark is projected via FC layers and reshaped to align with the dimensions of the hidden features.
Similarly positioned before the upsampling modules, WMAdapter~\cite{ci2025wmadapter} utilizes a Contextual Adapter to spatially align the watermark with the hidden features. To enhance the image quality, it employs a hybrid fine-tuning strategy to jointly train the Adapter and the Decoder.
StableGuard~\cite{yang2025stableguard} adheres to this embedding paradigm by projecting the watermark into a high-dimensional vector, which is subsequently reshaped to align with the spatial resolution of the hidden features.
Furthermore, we introduce two approaches leveraging Low-Rank Adaptation (LoRA) for watermarking. EW-LoRA similarly integrates LoRA modules into the LDM Decoder, optimizing the generative model via a specific objective function to directly synthesize watermarked images. 
In contrast, AquaLoRA targets the U-Net of the LDM, mapping the watermark into the latent space and superimposing it onto features.

\textbf{Challenge.}
While these methods offer robust protection for the visual field, inherent \textit{modal disparities} arising from the significantly lower redundancy and higher sensitivity render their direct transfer to speech.
Specifically, naively transposing the visual spatial tiling strategy to the temporal dimension induces perceptible periodic artifacts. 
This challenge is further compounded by vocoders operating on Mel-spectrograms rather than the latent space of LDM, which significantly narrows the margin for imperceptible embedding.
Furthermore, existing image approaches are specifically tailored to LDM, while speech synthesis is characterized by the widespread deployment of diverse vocoder architectures. 
This \textit{architectural diversity} necessitates a model-agnostic watermarking framework, distinguishing our approach from LDM-centric visual solutions.

\textbf{Solution.}
To address modal disparities, the TA employs \textit{Acoustic Feature Alignment} to embed watermarks into stable acoustic space, ensuring high perceptual speech generation.
This is complemented by \textit{Frame-level Temporal Broadcasting}, which eliminates the dependency on strict temporal synchronization, thereby robustly counteracting variable-length attacks.
Simultaneously, to overcome architectural diversity, our \textit{Progressive Feature Projection} dynamically adapts to the acoustic dimensions of disparate models, facilitating a unified framework across GAN-based and Diffusion-based vocoders.

\section{Model Architecture}
\subsection{Temporal Adapter}
\label{sec_app_ta_archi}
As illustrated in Fig.~\ref{fig_pipe}, the TA module comprises acoustic feature alignment, frame-level temporal broadcasting, and an adaptive injection mechanism. The detailed architecture is described as follows.

In the alignment mechanism, the Embedding layer consists of two FC layers and two LeakyReLU activations. 
The watermark $\mathbf{w}$ is first projected by the initial FC layer to a dimension equal to twice the hidden feature channels, followed by LeakyReLU activation. 
It is then mapped to $512$ dimensions via the second FC layer and processed by another LeakyReLU.

Subsequently, $Emb(\mathbf{w})$ enters the PFP, which contains two FC layers and a SiLU activation. 
Here, the first FC reduces the dimensionality to a pre-defined Rank value, followed by the SiLU, while the final FC aligns the output with the acoustic dimension of the hidden feature $h$.

Regarding the broadcasting mechanism, we employ the expand function to execute frame-level broadcasting. 
Finally, in the injection mechanism, the broadcasted features $\mathbf{w}_{\text{latent}}$ are initially concatenated with $h$ and then down-sampled via a 1-dimensional convolution layer (Conv1d). 
The sequence subsequently passes through a DSC and a zero-initialized Conv1d, with the final output $\hat{h}$ features obtained via a residual connection.

\subsection{Coarse-to-Fine Gated Extractor}
\label{sec_app_cage_archi}
As illustrated in Fig.~\ref{fig_pipe}, within the Coarse-to-Fine Feature Aggregation, each branch is constructed using a sequence of four GSCMs. 
Building upon the GSCM architecture detailed in Sec.~\ref{sec_ext}, we append Instance Normalization and LeakyReLU layers to further process the output, thereby enhancing convergence efficiency. 
The channel dimensions progress from $1$ to $32$, $64$, and $128$, with the final GSCM maintaining the channel count. 
The convolutional parameters vary by scale: $\mathrm{(i)}$ the fine-grained GSCM utilizes a kernel size of $3$, stride of $2$, and padding of $1$, $\mathrm{(ii)}$ the mid-grained scale uses a kernel size of $5$, stride of $2$, and padding of $2$, $\mathrm{(iii)}$ and the coarse-grained scale employs a kernel size of $7$, stride of $2$, and padding of $3$.

Subsequently, the output of each scale undergoes 1D adaptive average pooling. It is then processed by a GSCM ($k=1, s=1, p=0$) to reduce the channel dimension to $256$, followed by another GSCM ($k=3, s=1, p=1$) that maintains the dimensions. 
The resulting features are passed through parallel adaptive average pooling and adaptive max pooling layers, and the final representation is obtained via their arithmetic mean. Finally, this output is fed into a Decoder composed of two FC layers. 
The first projects the features to $512$ dimensions, and the second maps them to the final watermark length.

\begin{algorithm}[]
   \caption{Training of Accuracy-Guided Optimization Curriculum}
   \label{alg_training}
   \begin{algorithmic}[1]
      \item[] {\bfseries Input:} Mel-spectrogram $\mathbf{c}$, watermark $\mathbf{w}$, Frozen Generative Model with TA $\mathcal{G}(\cdot;\mathcal{T}_{\theta}(\cdot))$, Cage $\mathcal{E}_{\phi}(\cdot)$, Trainable parameters $\theta, \space \phi$ of TA and Cage, Attack Simulator $\mathcal{K}(\cdot)$, ACC Function $\mathcal{A}(\cdot)$, iteration steps $step$, trained epochs $Epoch$.
      
      \item[] {\bfseries Output:} TA $\mathcal{T}_{\theta}(\cdot)$, Cage $\mathcal{E}_{\phi}(\cdot)$
      
      \item[] Initialize hyper-parameters $\lambda_{\text{Mel}}=0.1, \space \lambda_{\text{mstft}}=0.1, \space \lambda_{\text{Ext}}=1$ 
      
      \FOR{$i=1$ {\bfseries to} $Epoch$}
         
         \FOR{$j=1$ {\bfseries to} $step$}
            \STATE Generate watermarked speech $\hat{\mathbf{s}} = \mathcal{G}(\mathbf{c}, \mathbf{w}; \theta)$.
            
            \STATE Extract watermark $\hat{\mathbf{w}} = \mathcal{E}_{\phi}(\mathcal{K}(\hat{\mathbf{s}}))$ under simulated attacks.
            
            \STATE Calculate batch accuracy $\text{acc}_j = \mathcal{A}(\mathbf{w}, \hat{\mathbf{w}})$.
            
            \STATE Compute total loss $\mathcal{L}$ via Eq.~\ref{eq_loss}.
            \STATE Update $\theta, \phi \leftarrow \text{AdamW}(\nabla \mathcal{L})$.
         \ENDFOR
         
         \STATE Calculate mean accuracy $\text{ACC} = \frac{1}{step}\sum_{j=1}^{step} \text{acc}_j$ over the current epoch.
         \STATE Update $\lambda_{\text{Mel}}, \space \lambda_{\text{mstft}}$ via Eq.~\ref{eq_agoc} according to $\text{ACC}$.
         
      \ENDFOR
   \end{algorithmic}
\end{algorithm}

\section{Accuracy-Guided Optimization Curriculum}
\label{sec_app_agoc}
Having detailed the training methodology of VocBulwark in Sec.~\ref{sec_train}, we further outline the specific training steps of the Accuracy-Guided Optimization Curriculum (AGOC) in Algorithm~\ref{alg_training}.

The training framework integrates the plug-in TA into the frozen generative model, synthesizing watermarked speech conditioned on the input Mel-spectrogram $\mathbf{c}$ and the unique watermark identifier $\mathbf{w}$. 
This generated speech is processed by Cage to recover the watermark under simulated attacks. 
Within the inner loop, the TA and Cage parameters are jointly optimized via gradient descent based on the total loss $\mathcal{L}$ by Eq.~\ref{eq_loss}, with weights initialized as $\lambda_{\text{Mel}}=\lambda_{\text{mstft}}=0.1$ and $\lambda_{\text{Ext}}=1$. 
Beyond the iteration steps, an outer feedback loop operates at the end of each epoch, dynamically recalibrating $\lambda_{\text{Mel}}$ and $\lambda_{\text{mstft}}$ according to the aggregated ACC and the curriculum strategy defined in Eq~\ref{eq_agoc}.

\section{Detailed Configurations of Diverse Attacks}
\label{sec_app_attack}
We categorize the robustness evaluation into four distinct classes: Common, Variable-length, Codec, and Compound attacks. 
To faithfully emulate real-world deployment scenarios characterized by unpredictable distortions, we focus on evaluating the \textit{robustness generalization} of our method. 
Specifically, we deliberately restricted the training stage to a subset of merely eight representative distortions with mild parameters. 
In contrast, the testing phase extends to a broader spectrum of \textit{unseen attacks} and employs significantly more aggressive parameter configurations for the seen types. 
This strict separation between training and testing conditions serves to rigorously validate the practical reliability of VocBulwark against unknown threats.

\subsection{Attacks for Training}
During training, we constructed an augmentation pool consisting of nine candidates: seven common post-processing operations, one Codec compression, and an identity operation. For each speech sample, we \textit{uniformly sampled} one operation from this pool to apply to the watermarked speech. 
The specific configurations are detailed as follows:
\begin{itemize}[itemsep=0pt, parsep=1pt, topsep=0pt]
    \item \textbf{Gaussian Noise:} Injects additive Gaussian noise with an intensity level uniformly sampled from $[15, 20]$.
    \item \textbf{Band-pass Filter:} Applies a band-pass filter retaining frequencies within the range of $1.5 \text{--} 10$ kHz.
    \item \textbf{High-pass Filter:} Applies a high-pass filter with a cutoff frequency threshold of $500$ Hz.
    \item \textbf{Sample Suppression:} Following \citet{chen2023wavmark}, randomly zeros out the initial, middle, or final segments of the waveform to simulate signal loss.
    \item \textbf{Dither:} Adds standard Triangular Probability Density Function (TPDF) dither to randomize quantization error.
    \item \textbf{Echo:} Introduces a synthetic acoustic echo effect to the speech signal.
    \item \textbf{Amplitude Scaling:} Scales the waveform amplitude by a random factor sampled from the range $[0.9, 1.1]$.
    \item \textbf{EnCodec:} Compresses and reconstructs the speech using the EnCodec neural audio tokenizer.
    \item \textbf{Identity:} Applies an identity mapping to represent the clean scenario.
\end{itemize}

\subsection{Common Attacks}
To comprehensively evaluate robustness to signal-processing distortions, we employ a diverse set of common attacks. The specific parameter configurations for the testing phase are detailed as follows:
\begin{itemize}[itemsep=0pt, parsep=1pt, topsep=0pt]
    \item \textbf{Gaussian Noise:} We add Gaussian Noise at varying intensities to test noise immunity. The evaluation covers a range of noise levels: $\{5, 10, 15, 20\}$ dB.
    \item \textbf{Low-Pass Filter (LPF):} Applies a low-pass filter with a cutoff frequency of $3$ kHz to simulate high-frequency loss.
    \item \textbf{Band-Pass Filter (BPF):} Applies a band-pass filter that retains spectral content solely within the range of $0.3 \text{--} 8$ kHz.
    \item \textbf{High-Pass Filter (HPF):} Applies a high-pass filter with a cutoff frequency of $1$ kHz.
    \item \textbf{Pink Noise:} Injects additive pink noise with the noise standard deviation set to 0.3.
    \item \textbf{Echo:} Introduces a synthetic echo effect using default parameters to simulate acoustic reverberation.
    \item \textbf{Dither:} Applies Triangular Probability Density Function (TPDF) dither to introduce randomized quantization error.
\end{itemize}

\subsection{Variable-length Attacks}
Variable-length attacks desynchronize the watermark by altering the temporal axis of the speech. We evaluate the model's resilience against the following temporal manipulations:
\begin{itemize}[itemsep=0pt, parsep=1pt, topsep=0pt]
    \item \textbf{Time Stretching:} The speech playback speed is modified without altering the pitch. We evaluate two scaling factors: $0.9\times$ (slowing down) and $1.1\times$ (speeding up).
    \item \textbf{Cropping:} We simulate signal loss at temporal boundaries by excising \textit{initial} or \textit{final} segments.
\end{itemize}

All post-processing operations are implemented using \texttt{torch} (version 1.10.1+cu111), \texttt{torchaudio} (version 0.10.1+cu111), and the AudioSeal~\cite{san2024audioseal}.

\subsection{Codec Attacks}
\label{sec_app_codec_config}
To assess robustness against quantization artifacts introduced by audio compression, we selected two representative Codecs ranging from classic to modern standards, as described in Sec.~\ref{sec_robust}. 
The detailed configurations are:
\begin{itemize}[itemsep=0pt, parsep=1pt, topsep=0pt]
    \item \textbf{MP3}~\cite{brandenburg1999mp3}\textbf{:} Utilizes the ubiquitous MP3 (MPEG-1 Audio Layer III) format for lossy compression and reconstruction. We evaluate performance at low-to-medium bitrates: $\{16, 32, 64\}$ kbps.
    \item \textbf{Opus}~\cite{valin2012opus}\textbf{:} Employs the modern Opus Codec, a versatile standard optimized for efficient speech transmission in the 5G era. Similarly, encoding and decoding are performed at bitrates of $\{16, 32, 64\}$ kbps.
\end{itemize}

The MP3 implementation relies on \texttt{audiomentations} (version 0.35.0), while Opus is derived from \texttt{opuslib} (version 3.0.1).

As outlined in Sec.~\ref{sec_robust}, we selected representative neural Codecs as robustness benchmarks, following \citet{Guo25discretespeech}. These Codecs cover three distinct categories: \textit{General-Purpose} Acoustic Token, \textit{Semantic Distillation} Acoustic Token, and \textit{Disentanglement} Acoustic Token:
\begin{itemize}[itemsep=0pt, parsep=1pt, topsep=0pt]
    \item \textbf{EnCodec} (General-Purpose)~\cite{defossez2023high}\textbf{:} A high-fidelity audio Codec leveraging residual vector quantization (RVQ). We employed bitrates of $\{3, 6, 12, 24\}$ kbps to attack the watermarked speech.

    \item \textbf{HiFi-Codec} (General-Purpose)~\cite{yang2023hificodec}\textbf{:} Utilizes the innovative Group-Residual Vector Quantization (GRVQ) mechanism to achieve high-fidelity audio reconstruction. We conduct compression and regeneration at a bitrate of $4$ kbps.

    \item \textbf{TiCodec} (General-Purpose)~\cite{ren2024ticodec}\textbf{:} Leverages a combination of RVQ and GVQ to extract time-invariant codes for efficient compression. We evaluate the regeneration performance at a bitrate of $3$ kbps.

    \item \textbf{SpeechTokenizer (STCodec)} (Semantic Distillation)~\cite{zhang2024speechtokenizer}\textbf{:} Built upon the RVQ, it employs semantic distillation to hierarchically disentangle semantic information from acoustic attributes. The speech is regenerated at a bitrate of $4$ kbps.

    \item \textbf{FACodec} (Disentanglement)~\cite{ju2024naturalspeech}\textbf{:} It employs a Factorized Vector Quantization (FVQ) to effectively disentangle speech into four independent subspaces of content, prosody, timbre, and acoustic details. In our experimental setup, the watermarked speech is regenerated at a bitrate of 4.8 kbps.
    
\end{itemize}

\subsection{Compound Attacks}
To simulate rigorous real-world transmission scenarios where speech often undergoes multi-stage processing, we evaluate robustness against compound attacks. These protocols involve the sequential application of two distinct distortion techniques. The specific combinations are defined as follows:
\begin{itemize}[itemsep=0pt, parsep=1pt, topsep=0pt]
    \item \textbf{GN+Ec:} Injection of Gaussian Noise followed by a synthetic acoustic Echo effect.
    \item \textbf{LPF+Dit:} Application of a Low-Pass Filter to truncate high frequencies, followed by TPDF Dither.
    \item \textbf{PN+GN:} Superposition of Pink Noise and Gaussian Noise to simulate complex, multi-source environmental background interference.
    \item \textbf{Dit+GN:} Application of TPDF Dither followed by additive Gaussian Noise.
    \item \textbf{Ec+Op:} Introduction of an acoustic Echo effect followed by lossy compression using the Opus Codec.
    \item \textbf{GN+Enc:} Injection of Gaussian Noise followed by neural re-synthesis via the EnCodec tokenizer.
    \item \textbf{Op+MP3:} A severe double-compression scenario where speech is first encoded with Opus and subsequently re-encoded using the MP3 standard.
    \item \textbf{Op+Enc:} Sequential compression involving traditional Opus encoding followed by neural regeneration via EnCodec, representing a hybrid transmission pipeline.
\end{itemize}

\section{More Details of Experimental Setting}
\label{sec_app_experiment}
In this section, we provide a detailed description of the datasets, evaluation metrics, and implementation details.

\subsection{Datasets}
We utilize the \textit{single-speaker English} dataset, LJSpeech~\cite{Ito2017ljspeech}, to train VocBulwark. 
To evaluate the genesimulates real-world deployment scenarios in whichto the \textit{multi-speaker English} LibriTTS~\cite{zen2019libritts} and the \textit{cross-lingual Chinese} AiShell3~\cite{shi2021aishell} datasets. 
This evaluation protocol is designed to simulate real-world deployment scenarios where target languages and speaker identities are often unpredictable, thereby validating the practical viability and robustness of our approach.
Regarding data preprocessing, all speech clips are segmented to 1 second and resampled to 22.05 kHz to align with the vocoder specifications.

\textbf{LJSpeech}~\cite{Ito2017ljspeech} is a single-speaker English corpus comprising 13,100 audio clips, each ranging from 1 to 10 seconds in duration. 
The data is derived from seven non-fiction books, totaling approximately 24 hours of speech recorded at a sampling rate of 22.05 kHz.

\textbf{LibriTTS}~\cite{zen2019libritts} is a multi-speaker English corpus derived from the original audio and text materials of LibriSpeech. 
Designed for text-to-speech tasks, it contains approximately 585 hours of speech from 2,456 speakers, sampled at 24 kHz.

\textbf{AiShell3}~\cite{shi2021aishell} is a multi-speaker Mandarin dataset designed for speech synthesis. 
It consists of 85 hours of recordings from 218 native speakers sampled at 44.1 kHz. 
The textual content covers domains such as smart home commands, news reports, and geographic information.

\begin{table}[t]
\centering
\caption{Quantitative Results of Robustness Against Common and Variable-length Attacks Under Large Capacity.}
\resizebox{0.95\textwidth}{!}{
\begin{tabular}{cccccccccccccccc}
\toprule
\multirow{3}{*}{Capacity} & \multirow{3}{*}{Method} & \multicolumn{10}{c}{Common Attacks}                                                           & \multicolumn{4}{c}{Variable-length Attacks}                     \\
\cmidrule(r){3-13}\cmidrule(r){13-16}
                          &                         & \multicolumn{4}{c}{Gaussian Noise} & LPF    & BPF    & HPF    & PN & Echo    & Dither & \multicolumn{2}{c}{Time Stretch} & \multicolumn{2}{c}{Cropping} \\
\cmidrule(r){3-6}\cmidrule(r){7-9}\cmidrule(r){10-10}\cmidrule(r){11-11}\cmidrule(r){12-12}\cmidrule(r){13-14}\cmidrule(r){15-16}
                          &                         & 5 dB    & 10 dB  & 15 dB  & 20 dB  & 3k     & 0.3-8k & 1k     & 0.3        & Default & TPDF   & 0.9$\times$     & 1.1$\times$    & Front         & Behind       \\
\midrule
\multicolumn{16}{c}{\textit{LJSpeech (In-Distribution Dataset)}}                \\
\midrule
\multirow{4}{*}{500 bps}  & VocBulwark[DW]            & 0.9334  & 0.9840 & 0.9931 & 0.9964 & 0.9981 & 0.8517 & 0.8545 & 0.8355     & 0.9974  & 0.9988 & 0.9998          & 0.9893         & 0.9836        & 0.9963       \\
                          & VocBulwark[PG]            & 0.9334  & 0.9841 & 0.9931 & 0.9964 & 0.9981 & 0.8517 & 0.8541 & 0.8353     & 0.9974  & 0.9988 & 0.9998          & 0.9893         & 0.9836        & 0.9963       \\
                          & VocBulwark[HFG]           & 0.8450  & 0.9680 & 0.9945 & 0.9997 & 1.0000 & 0.9946 & 0.9789 & 0.9636     & 0.9965  & 1.0000 & 0.9568          & 0.9527         & 0.9982        & 0.9997       \\
                          & VocBulwark[BVG]           & 0.9651  & 0.9974 & 0.9997 & 1.0000 & 1.0000 & 0.9080 & 0.9877 & 0.8298     & 0.9975  & 1.0000 & 0.9573          & 0.9496         & 0.9993        & 0.9993       \\
\midrule
\multirow{4}{*}{1000 bps} & VocBulwark[DW]            & 0.9575  & 0.9967 & 0.9988 & 0.9996 & 0.9522 & 0.9687 & 0.9988 & 0.8446     & 0.9996  & 0.9996 & 0.9912          & 0.9945         & 0.9893        & 0.9993       \\
                          & VocBulwark[PG]            & 0.9576  & 0.9967 & 0.9983 & 0.9996 & 0.9988 & 0.9522 & 0.9685 & 0.8445     & 0.9996  & 0.9996 & 0.9902          & 0.9942         & 0.9892        & 0.9993       \\
                          & VocBulwark[HFG]           & 0.8683  & 0.9365 & 0.9703 & 0.9835 & 0.9994 & 0.9995 & 0.9995 & 0.9982     & 0.9993  & 0.9995 & 0.9448          & 0.9022         & 0.9995        & 0.9990       \\
                          & VocBulwark[BVG]           & 0.8378  & 0.9320 & 0.9824 & 0.9965 & 0.9959 & 0.9785 & 0.9993 & 0.9993     & 0.9993  & 0.9993 & 0.9162          & 0.9137         & 0.9986        & 0.9965       \\
\midrule
\multirow{4}{*}{2000 bps} & VocBulwark[DW]            & 0.8638  & 0.9677 & 0.9954 & 0.9993 & 0.9993 & 0.8550 & 0.7965 & 0.8501     & 0.9998  & 0.9999 & 0.9990          & 0.9987         & 0.9827        & 0.9992       \\
                          & VocBulwark[PG]            & 0.8638  & 0.9676 & 0.9954 & 0.9993 & 0.9993 & 0.8550 & 0.7960 & 0.8489     & 0.9993  & 0.9993 & 0.9988          & 0.9986         & 0.9827        & 0.9990       \\
                          & VocBulwark[HFG]           & 0.8605  & 0.9328 & 0.9770 & 0.9928 & 0.9995 & 0.9985 & 0.9980 & 0.9993     & 0.9995  & 0.9993 & 0.9737          & 0.9873         & 0.9989        & 0.9983       \\
                          & VocBulwark[BVG]           & 0.8992  & 0.9678 & 0.9926 & 0.9987 & 0.9803 & 0.9065 & 1.0000 & 1.0000     & 1.0000  & 1.0000 & 0.9451          & 0.9882         & 1.0000        & 0.9988       \\
\midrule
\multicolumn{16}{c}{\textit{LibriTTS (OOD Dataset)}}               \\
\midrule
\multirow{4}{*}{500 bps}  & VocBulwark[DW]            & 0.9739  & 0.9815 & 0.9866 & 0.9910 & 0.9825 & 0.8504 & 0.8238 & 0.8606     & 0.9810  & 0.9919 & 0.9866          & 0.9935         & 0.9892        & 0.9905       \\
                          & VocBulwark[PG]            & 0.9730  & 0.9833 & 0.9890 & 0.9904 & 0.9877 & 0.8661 & 0.8707 & 0.8069     & 0.9838  & 0.9915 & 0.9872          & 0.9923         & 0.9824        & 0.9908       \\
                          & VocBulwark[HFG]           & 0.8160  & 0.9039 & 0.9626 & 0.9819 & 0.9989 & 0.9939 & 0.9684 & 0.8568     & 0.9775  & 0.9982 & 0.9872          & 0.9911         & 0.9938        & 0.9953       \\
                          & VocBulwark[BVG]           & 0.9009  & 0.9312 & 0.9505 & 0.9653 & 0.9072 & 0.9201 & 0.9013 & 0.9806     & 0.9952  & 0.9905 & 0.9820          & 0.9843         & 0.9831        & 0.9840       \\
\midrule
\multirow{4}{*}{1000 bps} & VocBulwark[DW]            & 0.9352  & 0.9748 & 0.9862 & 0.9894 & 0.9926 & 0.9905 & 0.9863 & 0.9304     & 0.9989  & 0.9929 & 0.9936          & 0.9916         & 0.9850        & 0.9947       \\
                          & VocBulwark[PG]            & 0.9353  & 0.9749 & 0.9894 & 0.9862 & 0.9926 & 0.9905 & 0.9865 & 0.9305     & 0.9989  & 0.9930 & 0.9936          & 0.9916         & 0.9849        & 0.9947       \\
                          & VocBulwark[HFG]           & 0.8627  & 0.9281 & 0.9667 & 0.9881 & 0.9979 & 0.9990 & 0.9990 & 0.9989     & 0.9990  & 0.9990 & 0.9698          & 0.9566         & 0.9988        & 0.9990       \\
                          & VocBulwark[BVG]           & 0.8816  & 0.9310 & 0.9597 & 0.9675 & 0.9705 & 0.9730 & 0.9899 & 0.9750     & 0.9856  & 0.9894 & 0.9631          & 0.9591         & 0.9841        & 0.9858       \\
\midrule
\multirow{4}{*}{2000 bps} & VocBulwark[DW]            & 0.8560  & 0.9189 & 0.9595 & 0.9768 & 0.9717 & 0.8709 & 0.8791 & 0.8476     & 0.9970  & 0.9756 & 0.9818          & 0.9828         & 0.9606        & 0.9778       \\
                          & VocBulwark[PG]            & 0.8562  & 0.9189 & 0.9595 & 0.9768 & 0.9717 & 0.8701 & 0.8792 & 0.8566     & 0.9971  & 0.9745 & 0.9820          & 0.9825         & 0.9606        & 0.9770       \\
                          & VocBulwark[HFG]           & 0.8513  & 0.9247 & 0.9660 & 0.9882 & 0.9980 & 0.9990 & 0.9989 & 0.9960     & 0.9990  & 0.9990 & 0.9494          & 0.9968         & 0.9960        & 0.9990       \\
                          & VocBulwark[BVG]           & 0.8147  & 0.8950 & 0.9318 & 0.9501 & 0.9671 & 0.9578 & 0.9970 & 0.9939     & 0.9939  & 0.9989 & 0.9085          & 0.9649         & 0.9919        & 0.9929       \\
\midrule
\multicolumn{16}{c}{\textit{AiShell3 (OOD Dataset)}}               \\
\midrule
\multirow{4}{*}{500 bps}  & VocBulwark[DW]            & 0.9262  & 0.9307 & 0.9344 & 0.9593 & 0.9299 & 0.8510 & 0.8904 & 0.8812     & 0.9032  & 0.9245 & 0.9287          & 0.9220         & 0.9569        & 0.8909       \\
                          & VocBulwark[PG]            & 0.9262  & 0.9305 & 0.9344 & 0.9593 & 0.9300 & 0.8511 & 0.8899 & 0.8812     & 0.9031  & 0.9246 & 0.9286          & 0.9220         & 0.9569        & 0.8910       \\
                          & VocBulwark[HFG]           & 0.8315  & 0.8921 & 0.9325 & 0.9460 & 0.9941 & 0.9651 & 0.9019 & 0.9860     & 0.9832  & 0.9930 & 0.9572          & 0.9241         & 0.9928        & 0.9865       \\
                          & VocBulwark[BVG]           & 0.8909  & 0.9212 & 0.9305 & 0.9362 & 0.8913 & 0.9401 & 0.9013 & 0.9156     & 0.9748  & 0.9805 & 0.9620          & 0.9243         & 0.9070        & 0.9218       \\
\midrule
\multirow{4}{*}{1000 bps} & VocBulwark[DW]            & 0.9600  & 0.9926 & 0.9933 & 0.9990 & 0.9979 & 0.9159 & 0.8194 & 0.9390     & 0.9947  & 0.9979 & 0.9989          & 0.9986         & 0.9926        & 0.9980       \\
                          & VocBulwark[PG]            & 0.9601  & 0.9926 & 0.9935 & 0.9988 & 0.9979 & 0.9115 & 0.8162 & 0.9388     & 0.9947  & 0.9979 & 0.9989          & 0.9979         & 0.9926        & 0.9982       \\
                          & VocBulwark[HFG]           & 0.8101  & 0.8593 & 0.9045 & 0.9565 & 0.9896 & 0.9988 & 0.9986 & 0.9412     & 0.9968  & 0.9989 & 0.9761          & 0.9931         & 0.9989        & 0.9893       \\
                          & VocBulwark[BVG]           & 0.7762  & 0.8232 & 0.9006 & 0.9409 & 0.9707 & 0.9680 & 0.9795 & 0.9340     & 0.9786  & 0.9915 & 0.9732          & 0.9526         & 0.9829        & 0.9861       \\
\midrule
\multirow{4}{*}{2000 bps} & VocBulwark[DW]            & 0.7934  & 0.8549 & 0.9172 & 0.9300 & 0.9294 & 0.8579 & 0.8577 & 0.8437     & 0.9814  & 0.9301 & 0.9380          & 0.9324         & 0.9138        & 0.9311       \\
                          & VocBulwark[PG]            & 0.7935  & 0.8547 & 0.9172 & 0.9302 & 0.9294 & 0.8581 & 0.8578 & 0.8440     & 0.9814  & 0.9303 & 0.9380          & 0.9323         & 0.9138        & 0.9312       \\
                          & VocBulwark[HFG]           & 0.7964  & 0.8601 & 0.9058 & 0.9503 & 0.9773 & 0.9919 & 0.9919 & 0.9240     & 0.9849  & 0.9921 & 0.9307          & 0.8729         & 0.9960        & 0.9868       \\
                          & VocBulwark[BVG]           & 0.7892  & 0.8405 & 0.9071 & 0.9291 & 0.7680 & 0.9586 & 0.9917 & 0.8978     & 0.9727  & 0.9745 & 0.9033          & 0.8798         & 0.9746        & 0.9839        \\
\bottomrule
\end{tabular}
}
\label{tab_app_std_atk}
\end{table}

\subsection{Definition of Evaluation Metrics}
As outlined in Sec.~\ref{sec_setting}, our evaluation metrics comprise five quantitative indicators across three domains. $\mathrm{(i)}$ Standard Objective Metrics (STOI~\cite{taal2010stoi} and PESQ~\cite{recommendation2001pesq}), $\mathrm{(ii)}$ Spectro-temporal Fidelity (SSIM~\cite{wang2004ssim} and MCD~\cite{kubichek1993mel}), and $\mathrm{(iii)}$ Neural Naturalness (DNSMOS). 
In the following, we provide detailed descriptions of these metrics.

\textbf{Short-Time Objective Intelligibility (STOI)}~\cite{taal2010stoi} quantifies speech intelligibility by calculating the spectro-temporal correlation between the clean reference and synthesized speech.
The resulting score is bounded within the range of $[0, 1]$, where a value closer to 1 indicates superior speech intelligibility.
The final metric is mathematically formulated as:
\begin{equation}
    \text{STOI} = \frac{1}{JM} \sum_{m=1}^{M} \sum_{j=1}^{J} d_j(m),
\end{equation}
where $M$ and $J$ denote the total number of frames and one-third octave bands, respectively. 
The term $d_j(m)$ represents the intermediate intelligibility measure, defined as the linear correlation coefficient of temporal envelopes for the $j$-th band at the $m$-th frame.

\textbf{Perceptual Evaluation of Speech Quality (PESQ)}~\cite{recommendation2001pesq}, standardized as ITU-T Recommendation P.862, is a widely adopted objective metric for assessing speech quality.
It operates by simulating the psychophysical properties of the human auditory system to quantify the perceptual difference between the clean reference and the degraded signal. The resulting score typically maps to a range of $[1.0, 4.5]$, where a higher value indicates superior perceptual quality closer to the original recording.

\textbf{Structural Similarity Index Measure (SSIM)}~\cite{wang2004ssim} evaluates the perceptual quality of the synthesized Mel-spectrograms by modeling distortion as a combination of correlation loss, luminance distortion, and contrast distortion.
By using the mean and variance as estimates of luminance and contrast, and the covariance as an estimate of structural similarity, SSIM quantifies how well the generated spectrogram preserves the fine-grained time-frequency patterns of the reference audio.
The structural similarity is quantitatively evaluated as:
\begin{equation}
    \text{SSIM}(\mathbf{x}, \mathbf{y}) = \frac{(2\mu_x \mu_y + C_1)(2\sigma_{xy} + C_2)}{(\mu_x^2 + \mu_y^2 + C_1)(\sigma_x^2 + \sigma_y^2 + C_2)},
\end{equation}
where $\mu_{(\cdot)}$ and $\sigma_{(\cdot)}^2$ represent the mean and variance of the spectral intensities, respectively, while $\sigma_{xy}$ denotes the covariance between the watermarked and original spectrograms. 
$C_1$ and $C_2$ act as stability constants to prevent division by zero.

\textbf{Mel-Cepstral Distortion (MCD)}~\cite{kubichek1993mel} is utilized to measure the spectral distance between the synthesized and reference audio. Based on our implementation, the metric is computed as the scaled Euclidean distance between the Mel-cepstral feature vectors. Mathematically, for a given pair of speech $\mathbf{c}_{wm}$ and $\mathbf{c}_{ori}$, the distortion is defined as:
\begin{equation}
    \text{MCD} = \frac{10\sqrt{2}}{\ln 10} \sqrt{\sum_{d=1}^{D} (c_{wm, d} - c_{ori, d})^2},
\end{equation}
where $c_{wm, d}$ and $c_{ori, d}$ denote the $d$-th coefficient of the watermarked and original speech, respectively. The scaling factor $\frac{10\sqrt{2}}{\ln 10}$ converts the metric into decibels (dB), ensuring consistency with standard acoustic evaluations.

\textbf{DNSMOS}~\cite{reddy2022dnsmos} is a non-intrusive perceptual objective speech quality metric developed based on the ITU-T P.835 subjective evaluation. 
Taking log power spectrograms as input, the Convolutional Neural Network-based model predicts three distinct scores: signal quality (SIG), background noise quality (BAK), and overall audio quality (OVRL). 
The metric is trained using a large-scale dataset from the 3rd Deep Noise Suppression (DNS) Challenge containing crowdsourced subjective ratings.
We use the SIG score as the definitive metric for assessing the quality of the watermarked speech.

\subsection{More Details of Implementation}
Complementing the training details outlined in Sec~\ref{sec_setting}, we explicitly restrict gradient updates to the TA and the Cage using the AdamW optimizer, while the pre-trained generative backbone remains frozen.
Furthermore, the loss balancing follows our Accuracy-Guided Optimization Curriculum.
Specifically, the perceptual weights are set to $\lambda_{\text{Mel}}=\lambda_{\text{mstft}}=0.2$ when $90\%<\text{ACC}<95\%$, and are increased to $\lambda_{\text{Mel}}=\lambda_{\text{mstft}}=0.5$ once $\text{ACC}>95\%$.
Moreover, we utilize the officially released code for all generative models without any modifications.
For the Diffusion Model, the TA is integrated into the final five residual layers of the denoising network, which is activated solely during the penultimate denoising step to ensure robustness while minimizing perceptual interference.
In the GAN architecture, the TA follows the first upsampling layer, where, facilitated by the Acoustic Feature Alignment mechanism, the TA's channel configuration is automatically scaled to synchronize with the varying feature map dimensions of the hierarchical upsampling network.

\begin{figure}[t]
    \centering
    \includegraphics[width=0.95\linewidth]{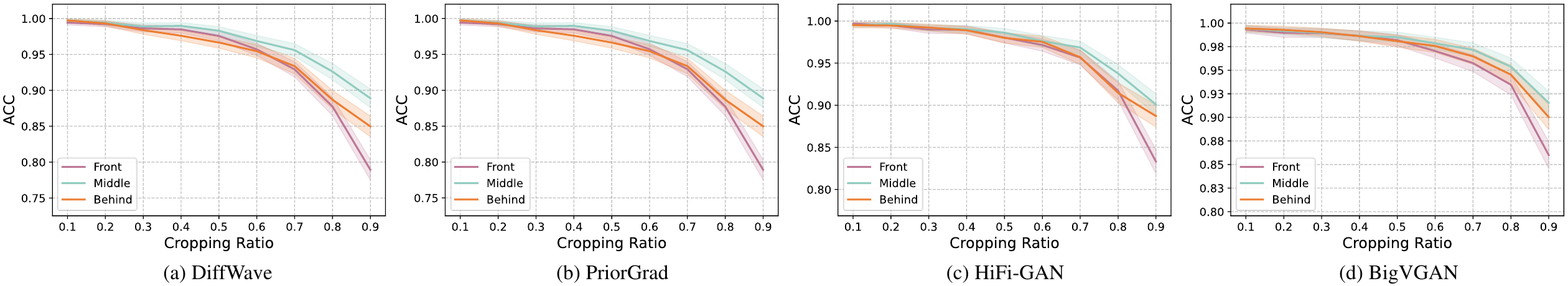}
    \caption{Robustness against cropping attacks across diverse ratios and positions.}
    \label{fig_app_crop}
\end{figure}

\begin{table*}[t]
\centering
\caption{Quantitative Results of Robustness Against Codec Compression and Regeneration Attacks Under Large Capacity.}
\resizebox{0.95\textwidth}{!}{
\begin{tabular}{cccccccccccccccc}
\toprule
&\multirow{3}{*}{} & \multicolumn{6}{c}{Traditional Codec Compression Attack}                     & \multicolumn{8}{c}{Neural Codec Regeneration Attack}                                               \\ 
\cmidrule(r){3-8}\cmidrule(r){9-16}
Capacity & Method (bps)   & \multicolumn{3}{c}{MP3}     & \multicolumn{3}{c}{Opus}    & \multicolumn{4}{c}{EnCodec}         & HiFiCodec & TiCodec & FACodec  & STCodec \\
\cmidrule(r){3-5}\cmidrule(r){6-8}\cmidrule(r){9-12}\cmidrule(r){13-14}\cmidrule(r){15-15}\cmidrule(r){16-16}
 & & 16 kbps & 32 kbps & 64 kbps & 16 kbps & 32 kbps & 64 kbps & 3 kbps & 6 kbps & 12 kbps & 24 kbps & 4 kbps    & 3 kbps  & 4.8 kbps & 4 kbps  \\
\midrule
\multicolumn{16}{c}{\textit{LJSpeech (In-Distribution Dataset)}}   \\
\midrule
\multirow{4}{*}{500 bps}  & VocBulwark[DW]      & 0.9967 & 0.9989 & 0.9992 & 0.9573 & 0.9634 & 0.9680 & 0.8980 & 0.9171 & 0.9567 & 0.9704 & 0.9019 & 0.8574 & 0.8888 & 0.9164 \\
                          & VocBulwark[PG]      & 0.9961 & 0.9985 & 0.9994 & 0.9560 & 0.9616 & 0.9700 & 0.8979 & 0.9147 & 0.9566 & 0.9700 & 0.9034 & 0.8572 & 0.8886 & 0.9125 \\
                          & VocBulwark[HFG]     & 0.9989 & 0.9989 & 0.9999 & 0.9989 & 0.9998 & 0.9999 & 0.9935 & 0.9992 & 0.9998 & 0.9998 & 0.9819 & 0.9558 & 0.9887 & 0.9813 \\
                          & VocBulwark[BVG]     & 0.9942 & 0.9985 & 1.0000 & 0.9913 & 0.9976 & 0.9993 & 0.9297 & 0.9903 & 0.9932 & 0.9992 & 0.8703 & 0.8982 & 0.9596 & 0.9181 \\
\midrule
\multirow{4}{*}{1000 bps} & VocBulwark[DW]      & 0.9993 & 0.9996 & 0.9996 & 0.9738 & 0.9786 & 0.9802 & 0.9601 & 0.9993 & 0.9996 & 0.9996 & 0.9993 & 0.9992 & 0.9606 & 0.9996 \\
                          & VocBulwark[PG]      & 0.9993 & 0.9996 & 0.9996 & 0.9765 & 0.9779 & 0.9789 & 0.9603 & 0.9993 & 0.9996 & 0.9996 & 0.9994 & 0.9996 & 0.9663 & 0.9998 \\
                          & VocBulwark[HFG]     & 0.9987 & 0.9993 & 0.9998 & 0.9995 & 0.9998 & 0.9998 & 0.9955 & 0.9988 & 0.9997 & 0.9997 & 0.9737 & 0.9922 & 0.9900 & 0.9908 \\
                          & VocBulwark[BVG]     & 0.9903 & 0.9931 & 0.9993 & 0.9745 & 0.9924 & 0.9945 & 0.8491 & 0.9214 & 0.9417 & 0.9418 & 0.8573 & 0.8714 & 0.8836 & 0.8927 \\
\midrule
\multirow{4}{*}{2000 bps} & VocBulwark[DW]      & 0.9954 & 0.9993 & 0.9993 & 0.8408 & 0.8453 & 0.8561 & 0.8353 & 0.8471 & 0.8556 & 0.9579 & 0.9915 & 0.9233 & 0.8793 & 0.9875 \\
                          & VocBulwark[PG]      & 0.9934 & 0.9993 & 0.9993 & 0.8416 & 0.8472 & 0.8569 & 0.8349 & 0.7471 & 0.8553 & 0.9576 & 0.9901 & 0.9258 & 0.9793 & 0.9863 \\
                          & VocBulwark[HFG]     & 0.9954 & 0.9993 & 0.9995 & 0.9884 & 0.9995 & 0.9995 & 0.8933 & 0.9567 & 0.9677 & 0.9760 & 0.8554 & 0.8647 & 0.8408 & 0.8891 \\
                          & VocBulwark[BVG]     & 0.8454 & 0.9010 & 0.9993 & 0.8471 & 0.8497 & 0.9294 & 0.8990 & 0.9542 & 0.9737 & 0.9782 & 0.7697 & 0.7995 & 0.8955 & 0.8825 \\
\midrule
\multicolumn{16}{c}{\textit{LibriTTS (OOD Dataset)}}      \\
\midrule
\multirow{4}{*}{500 bps}  & VocBulwark[DW]      & 0.9257 & 0.9891 & 0.9953 & 0.9299 & 0.9715 & 0.9862 & 0.8398 & 0.9288 & 0.9268 & 0.9477 & 0.9689 & 0.9413 & 0.9778 & 0.9031 \\
                          & VocBulwark[PG]      & 0.9260 & 0.9846 & 0.9915 & 0.9308 & 0.9726 & 0.9871 & 0.8705 & 0.9270 & 0.9569 & 0.9553 & 0.9631 & 0.9402 & 0.9346 & 0.9442 \\
                          & VocBulwark[HFG]     & 0.9938 & 0.9970 & 0.9982 & 0.9964 & 0.9956 & 0.9973 & 0.9142 & 0.9878 & 0.9934 & 0.9957 & 0.9642 & 0.8779 & 0.9006 & 0.9230 \\
                          & VocBulwark[BVG]     & 0.8786 & 0.8914 & 0.9045 & 0.8722 & 0.8796 & 0.9023 & 0.8757 & 0.9346 & 0.9360 & 0.9413 & 0.8707 & 0.8622 & 0.9152 & 0.9386 \\
\midrule
\multirow{4}{*}{1000 bps} & VocBulwark[DW]      & 0.9915 & 0.9926 & 0.9968 & 0.9300 & 0.9389 & 0.9405 & 0.9837 & 0.9917 & 0.9958 & 0.9960 & 0.9936 & 0.9894 & 0.9540 & 0.9935 \\
                          & VocBulwark[PG]      & 0.9912 & 0.9919 & 0.9995 & 0.9288 & 0.9393 & 0.9401 & 0.9837 & 0.9926 & 0.9958 & 0.9958 & 0.9926 & 0.9894 & 0.9524 & 0.9926 \\
                          & VocBulwark[HFG]     & 0.9713 & 0.9990 & 0.9990 & 0.9981 & 0.9990 & 0.9991 & 0.9596 & 0.9902 & 0.9915 & 0.9925 & 0.9201 & 0.9766 & 0.9672 & 0.9970 \\
                          & VocBulwark[BVG]     & 0.9674 & 0.9703 & 0.9914 & 0.9358 & 0.9650 & 0.9820 & 0.8008 & 0.8855 & 0.9186 & 0.9162 & 0.8551 & 0.8521 & 0.8803 & 0.8705 \\
\midrule
\multirow{4}{*}{2000 bps} & VocBulwark[DW]      & 0.9778 & 0.9786 & 0.9808 & 0.8651 & 0.8665 & 0.8707 & 0.7690 & 0.8133 & 0.8415 & 0.8496 & 0.9679 & 0.9408 & 0.8129 & 0.9728 \\
                          & VocBulwark[PG]      & 0.9751 & 0.9788 & 0.9788 & 0.8643 & 0.8664 & 0.8696 & 0.7708 & 0.8123 & 0.8446 & 0.8527 & 0.9681 & 0.9414 & 0.8063 & 0.9715 \\
                          & VocBulwark[HFG]     & 0.9939 & 0.9980 & 0.9990 & 0.9849 & 0.9990 & 0.9990 & 0.8804 & 0.9368 & 0.9491 & 0.9504 & 0.7732 & 0.8039 & 0.8173 & 0.8788 \\
                          & VocBulwark[BVG]     & 0.8387 & 0.8941 & 0.9958 & 0.8548 & 0.9498 & 0.9518 & 0.8002 & 0.8422 & 0.8683 & 0.8702 & 0.8721 & 0.8690 & 0.8214 & 0.8527 \\
\midrule
\multicolumn{16}{c}{\textit{AiShell3 (OOD Dataset)}}      \\
\midrule
\multirow{4}{*}{500 bps}  & VocBulwark[DW]      & 0.9165 & 0.9297 & 0.9286 & 0.9146 & 0.9247 & 0.9339 & 0.8305 & 0.8916 & 0.9380 & 0.9672 & 0.8876 & 0.8743 & 0.8097 & 0.8726 \\
                          & VocBulwark[PG]      & 0.9168 & 0.9293 & 0.9303 & 0.9141 & 0.9244 & 0.9329 & 0.8308 & 0.8926 & 0.9392 & 0.9683 & 0.8863 & 0.8742 & 0.8090 & 0.8745 \\
                          & VocBulwark[HFG]     & 0.9891 & 0.9946 & 0.9967 & 0.9890 & 0.9950 & 0.9945 & 0.8837 & 0.9604 & 0.9820 & 0.9908 & 0.9014 & 0.8142 & 0.8660 & 0.9157 \\
                          & VocBulwark[BVG]     & 0.8600 & 0.8726 & 0.8839 & 0.8563 & 0.8925 & 0.8780 & 0.7918 & 0.8270 & 0.8240 & 0.8320 & 0.8518 & 0.8282 & 0.8990 & 0.8993 \\
\midrule
\multirow{4}{*}{1000 bps} & VocBulwark[DW]      & 0.9968 & 0.9979 & 0.9989 & 0.9103 & 0.9278 & 0.9279 & 0.9979 & 0.9989 & 0.9989 & 0.9989 & 0.9926 & 0.9989 & 0.9667 & 0.9896 \\
                          & VocBulwark[PG]      & 0.9968 & 0.9968 & 0.9989 & 0.9114 & 0.9270 & 0.9289 & 0.9979 & 0.9982 & 0.9989 & 0.9989 & 0.9915 & 0.9989 & 0.9689 & 0.9903 \\
                          & VocBulwark[HFG]     & 0.9220 & 0.9958 & 1.0000 & 0.9378 & 0.9928 & 0.9989 & 0.8038 & 0.8254 & 0.8405 & 0.8507 & 0.7730 & 0.7762 & 0.8147 & 0.8399 \\
                          & VocBulwark[BVG]     & 0.9603 & 0.9747 & 0.9873 & 0.9446 & 0.9605 & 0.9697 & 0.7859 & 0.8357 & 0.8561 & 0.8672 & 0.8423 & 0.8677 & 0.8774 & 0.8587 \\
\midrule
\multirow{4}{*}{2000 bps} & VocBulwark[DW]      & 0.9273 & 0.9356 & 0.9475 & 0.8773 & 0.8850 & 0.8859 & 0.8660 & 0.8827 & 0.9060 & 0.9121 & 0.8885 & 0.9112 & 0.8101 & 0.9450 \\
                          & VocBulwark[PG]      & 0.9311 & 0.9364 & 0.9475 & 0.8785 & 0.8813 & 0.8842 & 0.8656 & 0.8865 & 0.9071 & 0.9113 & 0.8847 & 0.9163 & 0.8125 & 0.9433 \\
                          & VocBulwark[HFG]     & 0.9859 & 0.9879 & 0.9903 & 0.9247 & 0.9758 & 0.9871 & 0.8120 & 0.8144 & 0.8276 & 0.8464 & 0.7694 & 0.7927 & 0.7962 & 0.8464 \\
                          & VocBulwark[BVG]     & 0.8557 & 0.9050 & 0.9718 & 0.8438 & 0.8498 & 0.8558 & 0.7728 & 0.8211 & 0.8637 & 0.8589 & 0.7485 & 0.7637 & 0.8202 & 0.8247 \\ 
\bottomrule
\end{tabular}
}
\label{tab_app_Codec_atk}
\end{table*}

\begin{table}[t]
\centering
\caption{Quantitative Results of Robustness Against Compound Attacks Under Large Capacity.}
\resizebox{0.7\textwidth}{!}{
\begin{tabular}{cccccccccc}
\toprule
Capacity                  & Method (bps)  & GN+Ec  & LPF+Dit & PN+GN  & Dit+GN & Ec+Op  & GN+Enc & Op+MP3 & Op+Enc \\ 
\midrule
\multicolumn{10}{c}{\textit{LJSpeech (In-Distribution Dataset)}}                                                                       \\
\midrule
\multirow{4}{*}{500 bps}  & VocBulwark[DW]  & 0.9990 & 0.9981  & 0.8455 & 0.9982 & 0.9980 & 0.9475 & 0.9634 & 0.8617 \\
                          & VocBulwark[PG]  & 0.9989 & 0.9981  & 0.8446 & 0.9980 & 0.9972 & 0.9459 & 0.9657 & 0.8708 \\
                          & VocBulwark[HFG] & 0.9790 & 0.9999  & 0.8873 & 0.9948 & 0.9890 & 0.9766 & 0.9999 & 0.9996 \\
                          & VocBulwark[BVG] & 0.9929 & 1.0000  & 0.8953 & 0.9985 & 0.9767 & 0.9905 & 0.9970 & 0.9981 \\
\midrule
\multirow{4}{*}{1000 bps} & VocBulwark[DW]  & 0.9996 & 0.9992  & 0.8257 & 0.9986 & 0.9996 & 0.9986 & 0.9717 & 0.9620 \\
                          & VocBulwark[PG]  & 0.9990 & 0.9988  & 0.8258 & 0.9990 & 0.9991 & 0.9987 & 0.9682 & 0.9638 \\
                          & VocBulwark[HFG] & 0.9989 & 0.9989  & 0.9748 & 0.8585 & 0.9958 & 0.8207 & 0.9965 & 0.9840 \\
                          & VocBulwark[BVG] & 0.9789 & 0.9965  & 0.9732 & 0.9553 & 0.9924 & 0.8051 & 0.9920 & 0.9075 \\
\midrule
\multirow{4}{*}{2000 bps} & VocBulwark[DW]  & 0.9993 & 0.9973  & 0.8207 & 0.9918 & 0.9993 & 0.8442 & 0.8390 & 0.8533 \\
                          & VocBulwark[PG]  & 0.9989 & 0.9968  & 0.8216 & 0.9918 & 0.9990 & 0.7414 & 0.8399 & 0.7519 \\
                          & VocBulwark[HFG] & 0.8658 & 0.9987  & 0.9567 & 0.9128 & 0.9932 & 0.8521 & 0.9961 & 0.9615 \\
                          & VocBulwark[BVG] & 0.9747 & 0.8854  & 0.9851 & 0.8612 & 0.8095 & 0.8459 & 0.8504 & 0.8366 \\
\midrule
\multicolumn{10}{c}{\textit{LibriTTS (OOD Dataset)}}                                                                     \\
\midrule
\multirow{4}{*}{500 bps}  & VocBulwark[DW]  & 0.9858 & 0.9843  & 0.8010 & 0.9831 & 0.9730 & 0.9274 & 0.8974 & 0.8737 \\
                          & VocBulwark[PG]  & 0.9849 & 0.9915  & 0.8060 & 0.9871 & 0.9735 & 0.9317 & 0.9291 & 0.8780 \\
                          & VocBulwark[HFG] & 0.9093 & 0.9988  & 0.7799 & 0.9616 & 0.9680 & 0.9220 & 0.9965 & 0.9915 \\
                          & VocBulwark[BVG] & 0.9003 & 0.9792  & 0.8494 & 0.9491 & 0.9162 & 0.9362 & 0.9686 & 0.9406 \\
\midrule
\multirow{4}{*}{1000 bps} & VocBulwark[DW]  & 0.9995 & 0.9913  & 0.8444 & 0.9872 & 0.9996 & 0.9894 & 0.9345 & 0.9365 \\
                          & VocBulwark[PG]  & 0.9995 & 0.9914  & 0.8445 & 0.9872 & 0.9992 & 0.9913 & 0.9383 & 0.9361 \\
                          & VocBulwark[HFG] & 0.8482 & 0.9979  & 0.9365 & 0.8288 & 0.9989 & 0.8424 & 0.9989 & 0.8351 \\
                          & VocBulwark[BVG] & 0.9137 & 0.9725  & 0.9008 & 0.9014 & 0.9545 & 0.7964 & 0.9462 & 0.8392 \\
\midrule
\multirow{4}{*}{2000 bps} & VocBulwark[DW]  & 0.9994 & 0.9718  & 0.8572 & 0.9684 & 0.9970 & 0.8418 & 0.8773 & 0.7802 \\
                          & VocBulwark[PG]  & 0.9993 & 0.9718  & 0.8569 & 0.9685 & 0.9967 & 0.8409 & 0.8715 & 0.7832 \\
                          & VocBulwark[HFG] & 0.9008 & 0.9970  & 0.9280 & 0.8745 & 0.9970 & 0.8220 & 0.9960 & 0.9262 \\
                          & VocBulwark[BVG] & 0.9015 & 0.8558  & 0.9037 & 0.8368 & 0.8870 & 0.7912 & 0.8477 & 0.8356 \\
\midrule
\multicolumn{10}{c}{\textit{AiShell3 (OOD Dataset)}}                                                                     \\
\midrule
\multirow{4}{*}{500 bps}  & VocBulwark[DW]  & 0.9122 & 0.9273  & 0.8555 & 0.9253 & 0.8813 & 0.8868 & 0.8242 & 0.8230 \\
                          & VocBulwark[PG]  & 0.9126 & 0.9278  & 0.8552 & 0.9252 & 0.8810 & 0.8894 & 0.8314 & 0.8132 \\
                          & VocBulwark[HFG] & 0.8632 & 0.9957  & 0.8049 & 0.9483 & 0.9655 & 0.8367 & 0.9873 & 0.9846 \\
                          & VocBulwark[BVG] & 0.7913 & 0.8829  & 0.8140 & 0.8225 & 0.8399 & 0.8070 & 0.8707 & 0.8384 \\
\midrule
\multirow{4}{*}{1000 bps} & VocBulwark[DW]  & 0.9979 & 0.9969  & 0.8489 & 0.9936 & 0.9995 & 0.9979 & 0.9269 & 0.9777 \\
                          & VocBulwark[PG]  & 0.9979 & 0.9970  & 0.8491 & 0.9936 & 0.9993 & 0.9988 & 0.9267 & 0.9787 \\
                          & VocBulwark[HFG] & 0.8804 & 0.9768  & 0.7838 & 0.8629 & 0.9589 & 0.8023 & 0.9489 & 0.8430 \\
                          & VocBulwark[BVG] & 0.8618 & 0.9958  & 0.7859 & 0.8721 & 0.9907 & 0.7767 & 0.9841 & 0.8207 \\
\midrule
\multirow{4}{*}{2000 bps} & VocBulwark[DW]  & 0.9992 & 0.9355  & 0.7573 & 0.8939 & 0.9770 & 0.7912 & 0.7901 & 0.8619 \\
                          & VocBulwark[PG]  & 0.9988 & 0.9356  & 0.7568 & 0.8942 & 0.9745 & 0.7871 & 0.7976 & 0.8547 \\
                          & VocBulwark[HFG] & 0.8649 & 0.9778  & 0.7765 & 0.7946 & 0.9756 & 0.7939 & 0.9606 & 0.8168 \\
                          & VocBulwark[BVG] & 0.8454 & 0.8624  & 0.8752 & 0.8465 & 0.9120 & 0.8709 & 0.8498 & 0.8327 \\
\bottomrule
\end{tabular}
}
\label{tab_app_cmp_atk}
\end{table}

\section{More Experimental Results and Analysis}
\label{sec_app_exp_analysis}
In this section, we supplement the experimental results and analysis in Sec.~\ref{sec_robust} and Sec.~\ref{sec_eff} by scrutinizing the system from a multi-dimensional perspective, covering its robustness, computational complexity, and generalization.

\subsection{Robustness Against Standard Attacks}
\label{sec_app_standard_atk}
Following the evaluation detailed in Sec.~\ref{sec_robust}, Table~\ref{tab_app_std_atk} summarizes the robustness on the in-distribution LJSpeech dataset under high-capacity settings. 
Attributed to the Frame-level Temporal Broadcasting, VocBulwark exhibits universal stability against variable-length attacks. 
Specifically, Diffusion-based vocoders achieve superior accuracies exceeding 98\% against TS and Cropping, establishing a robust trend consistently observed across GAN-based architectures even as capacity scales.
Regarding signal degradations, the method maintains robust acoustic entanglement, sustaining resilient performance under noisy scenarios.
Even under severe Gaussian noise interference at 5 dB, the Diffusion and GAN-based models retain accuracies of approximately 93\% and 90\%, respectively. 
Furthermore, this robustness persists across diverse filtering conditions and expanded watermark capacities, collectively underscoring the method's scalability and reliability in real-world environments.

In addition, to further validate \textit{Frame-level Temporal Broadcasting}, we evaluated robustness against cropping attacks, as shown in Fig.~\ref{fig_app_crop}.
Although cropping attacks inevitably disrupt the synchronization required for watermark extraction, the experimental results provide compelling evidence that our broadcasting mechanism effectively eliminates temporal-domain dependence, thereby conferring stable resistance against diverse cropping scenarios.
Specifically, for Diffusion-based models, our VocBulwark maintains an accuracy surpassing 90\% even at a substantial cropping ratio of 0.7. 
Remarkably, the GAN-based variants exhibit superior endurance, with our approach maintaining over 90\% accuracy even when the cropping ratio reaches 0.8. 
Furthermore, the narrow confidence intervals indicate minimal performance variance across different samples, highlighting the stability and reliability of our method regardless of the cropping position.

\subsection{Robustness Against Codec Attacks}
\label{sec_app_codec_atk}
Consistent with the experimental configuration in Sec.~\ref{sec_robust}, we further evaluated robustness against Codec attacks under increased watermark capacities, as illustrated in Table~\ref{tab_app_Codec_atk}.
Unlike traditional Codecs, which primarily operate at the signal level, VocBulwark leverages acoustic space embedding to effectively counteract quantization artifacts, enabling our variants to maintain high accuracy across varying capacities. 
This is exemplified by our DW, which sustains ACC exceeding 98\% under MP3 and Opus compression in 64 kbps even at a high capacity of 2000 bps. 
In terms of Neural Codec regeneration, our method further enhances robustness through acoustic property embedding and a proposed optimization curriculum that addresses the challenges of vector quantization and extreme compression rates. 
Consequently, VocBulwark achieves exceptional performance against general-purpose neural Codecs, with our DW attaining 99\% ACC under EnCodec at 1000 bps. 
While confronting semantic distillation and disentanglement-based Codecs like STCCodec and FACodec inevitably incur greater information loss, leading to a marginal performance decline, our method retains substantial robustness, as HFG maintains a commendable ACC of about 98\%.

\subsection{Robustness Against Compound Attacks}
\label{sec_app_cmp_atk}
Consistent with the experimental configuration in Sec.~\ref{sec_robust}, we evaluated robustness against compound attacks under high-capacity conditions. 
The results in Table~\ref{tab_app_cmp_atk} demonstrate that our method maintains exceptional stability even when diverse signal degradations are superimposed. 
Specifically, when subjected to combinations of two common attacks, our Diffusion-based and GAN-based backbones achieve average accuracies of 96.51\% and 96.85\% at 500 bps, respectively, establishing a resistance that remains substantial as capacity increases. 
Furthermore, in hybrid scenarios involving both common and Codec compression, VocBulwark consistently withstands the combined interference. Although compound attacks comprising dual Codec impose significantly higher destructiveness, our method successfully recovers the watermark from the attacked speech. 
This is evidenced by DW and PG retaining an average accuracy exceeding 90\%, while HFG and BVG achieve remarkable performance reaching 99\%, confirming that VocBulwark possesses reliable extraction capabilities against complex distortions even as payload capacity scales.

\subsection{Computational Efficiency Analysis}
\label{sec_app_compute}
Extending the efficiency analysis from Sec.~\ref{sec_eff}, we present a granular breakdown of model sizes encompassing the Vocoder, Watermark Encoder (Adapter), and Extractor alongside a detailed FLOPs evaluation for both generation and extraction processes, as illustrated in Table~\ref{tab_app_complexity}.
Specifically, for Diffusion-based architectures, VocBulwark demonstrates distinct structural advantages, requiring only single-digit megabytes of storage. 
This stands in stark contrast to RIWF, which demands exceeding $7\times$ this capacity, and Groot, whose encoder imposes an excessive memory burden surpassing ours by $237\times$. 
While generation costs remain comparable across methods, our extraction process strikes a favorable balance by being significantly more efficient than RIWF. 
Such compactness is equally evident in GAN-based vocoders, where our lightweight decoder is approximately $26\times$ smaller than the HiFi-GANw counterpart. Notably, while generation FLOPs show negligible differences, our extraction overhead is reduced by over $10\times$ to ensure rapid retrieval suitable for real-time applications.

\begin{figure}[t]
    \centering
    \includegraphics[width=0.7\linewidth]{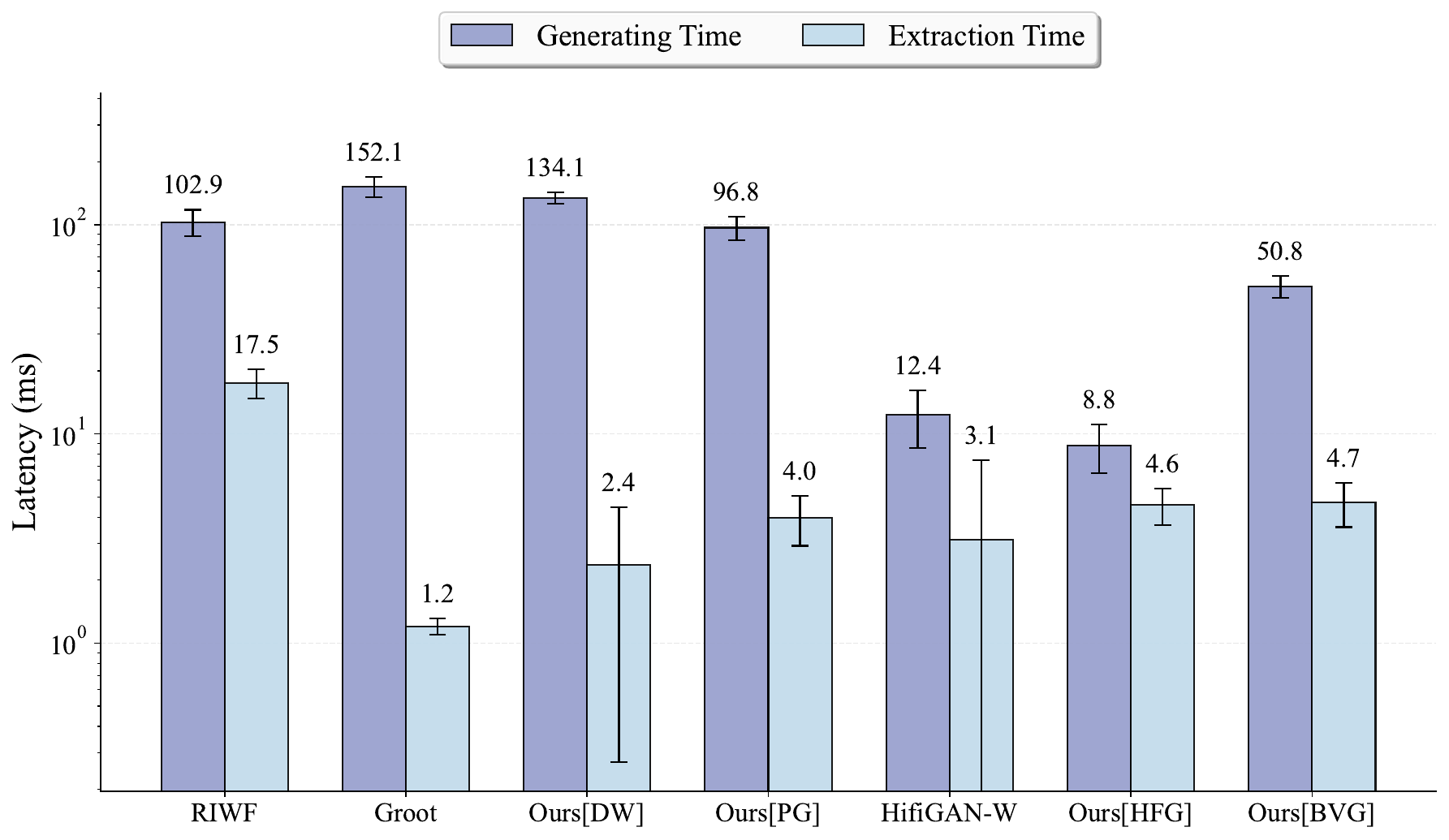}
    \caption{Comparison of Computational Efficiency in Generating and Extraction Time.}
    \label{fig_app_infer_time}
\end{figure}

To further assess real-time applicability, we visualized the inference latency for both generation and extraction in Fig.~\ref{fig_app_infer_time}. 
Benefiting from the lightweight nature of \textit{Additional-Parameter Injection}, VocBulwark demonstrates remarkable efficiency with negligible computational overhead.
In the realm of diffusion models, while generation latency is inherently dominated by iterative denoising, our streamlined extraction process significantly optimizes performance. 
Specifically, our PG demonstrates superior generation speed, while our DW drastically reduces extraction latency to 2.4 ms, approximately $7\times$ faster than RIWF.
This efficiency advantage extends to GAN-based architectures, where our HFG achieves a generation speed surpassing that of the HiFi-GANw baseline (8.8 ms vs. 12.4 ms).
Crucially, empirical observations confirm that this embedding strategy imposes negligible latency overhead on the original generation process, ensuring that the high-speed synthesis capability of the backbone model remains intact.

\begin{table}[t]
\centering
\caption{Detailed Comparison of Computational Complexity.}
\resizebox{0.8\textwidth}{!}{
\begin{tabular}{cccccccccc}
\toprule
\multirow{2}{*}{Method (bps)}                                    & \multicolumn{2}{c}{Training Params.~$\downarrow$} & \multicolumn{4}{c}{Model Size (MB)~$\downarrow$} & \multicolumn{3}{c}{FLOPs~$\downarrow$} \\
\cmidrule(r){2-3}\cmidrule(r){4-7}\cmidrule(r){8-10}

 & Stage 1                    & Stage 2                   & Embedding      & Extracting    & Vocoder   & Total   & Generating    & Extracting   & Total        \\
\midrule
\multicolumn{10}{c}{\textit{Diffusion-Based Vocoder}} \\
\midrule
RIWF (16)                                                        & 0.03 M                     & 4.43M                     & 20.36    & 16.63         & 30.22     & 67.21  & 3.44$\times 10^{11}$    & 8.34$\times 10^{9}$    & 3.52$\times 10^{11}$   \\
Groot (100)                                                      & 250.28 M                   & -                         & 866.34         & 88.43         & 30.22   & 984.99   & 3.48$\times 10^{11}$    & 0.40$\times 10^{9}$    & 3.49$\times 10^{11}$   \\
\rowcolor[HTML]{f7fcfd} 
Ours[DW] (100)            & 1.66 M                     & -                         & 3.73           & 2.70          & 30.22     & 40.06    & 3.57$\times 10^{11}$    & 1.66$\times 10^{9}$    & 3.59$\times 10^{11}$   \\
\rowcolor[HTML]{f7fcfd} 
Ours[PG] (100)            & 1.66 M                     & -                         & 3.73           & 2.70          & 30.22     & 40.06    & 3.57$\times 10^{11}$    & 1.66$\times 10^{9}$    & 3.59$\times 10^{11}$   \\
\midrule
\multicolumn{10}{c}{\textit{GAN-Based Vocoder}} \\
\midrule
HiFi-GANw (20)                                                   & 0.77 M*                    & 13.94 M                   & -              & 70.70         & 53.20     & 123.90  & 6.09$\times 10^{10}$    & 2.65$\times 10^{10}$   & 8.75$\times 10^{10}$   \\
\rowcolor[HTML]{f7fcfd} 
Ours[HFG] (100)           & 1.76 M                     & -                         & 4.09           & 2.70          & 53.20     & 59.99     & 6.18$\times 10^{10}$    & 1.66$\times 10^{9}$    & 6.35$\times 10^{10}$   \\
\rowcolor[HTML]{f7fcfd} 
Ours[BVG] (100)           & 1.76 M                     & -                         & 4.09           & 2.70          & 53.42     & 60.21    & 6.18$\times 10^{10}$    & 1.66$\times 10^{9}$    & 6.35$\times 10^{10}$   \\
\bottomrule
\end{tabular}
}
\label{tab_app_complexity}
\end{table}

\begin{table}[t]
\centering
\scriptsize
\caption{Generalization of Fidelity on Out-of-Distribution LibriTTS and AiShell3 Dataset at 100 bps.}
\resizebox{0.7\linewidth}{!}{
\begin{tabular}{ccccccc}
\toprule
Method   & STOI~$\uparrow$   & PESQ~$\uparrow$   & SSIM~$\uparrow$   & MCD~$\downarrow$     & DNSMOS~$\uparrow$ & ACC~$\uparrow$    \\ 
\midrule
\multicolumn{7}{c}{\textit{LibriTTS (OOD Dataset)}}   \\
\midrule
DiffWave       & 0.9232 & 2.6435 & 0.7924 & 5.2536  & 2.8786 &        \\
VocBulwark[DW]       & 0.9097 & 2.5057 & 0.7998 & 5.2630  & 2.9453 & 0.9987 \\
PriorGrad      & 0.9224 & 2.6454 & 0.7924 & 5.2651  & 2.8823 &        \\
VocBulwark[PG]       & 0.9097 & 2.5057 & 0.7998 & 5.2634  & 2.9452 & 0.9987 \\
HiFi-GAN       & 0.9365 & 2.4212 & 0.8839 & 4.3835  & 2.8813 &        \\
VocBulwark[HFG]     & 0.9352 & 2.2122 & 0.8557 & 4.2222  & 3.0477 & 0.9954 \\
BigVGAN        & 0.9789 & 3.6240 & 0.9382 & 4.0743  & 3.1664 & -      \\
VocBulwark[BVG]      & 0.9765 & 3.5814 & 0.9326 & 3.9814  & 3.3317 & 0.9919 \\ 
\midrule
\multicolumn{7}{c}{\textit{AiShell3 (OOD Dataset)}}   \\
\midrule
DiffWave       & 0.8909 & 2.3236 & 0.5165 & 8.7187  & 2.6128 & -      \\
VocBulwark[DW]       & 0.8786 & 2.1895 & 0.5312 & 9.5416  & 2.5684 & 0.9880 \\
PriorGrad      & 0.8919 & 2.3297 & 0.5164 & 8.7115  & 2.6198 & -      \\
VocBulwark[PG       & 0.8786 & 2.1894 & 0.5312 & 9.5419  & 2.5685 & 0.9880 \\
HiFi-GAN       & 0.9219 & 2.5935 & 0.7451 & 4.6982  & 2.7447 & -      \\
VocBulwark[HFG]      & 0.9089 & 2.1657 & 0.6448 & 4.6220  & 2.8196 & 0.9824 \\
BigVGAN        & 0.9690 & 3.4614 & 0.9033 & 4.8948  & 2.9512 & -      \\
VocBulwark[BVG]      & 0.9678 & 3.4325 & 0.9338 & 2.9047  & 3.1087 & 0.9826 \\ 
\bottomrule
\end{tabular}
}
\label{tab_app_fidelity}
\end{table}

\begin{figure}[]
    \centering
    \includegraphics[width=0.95\textwidth]{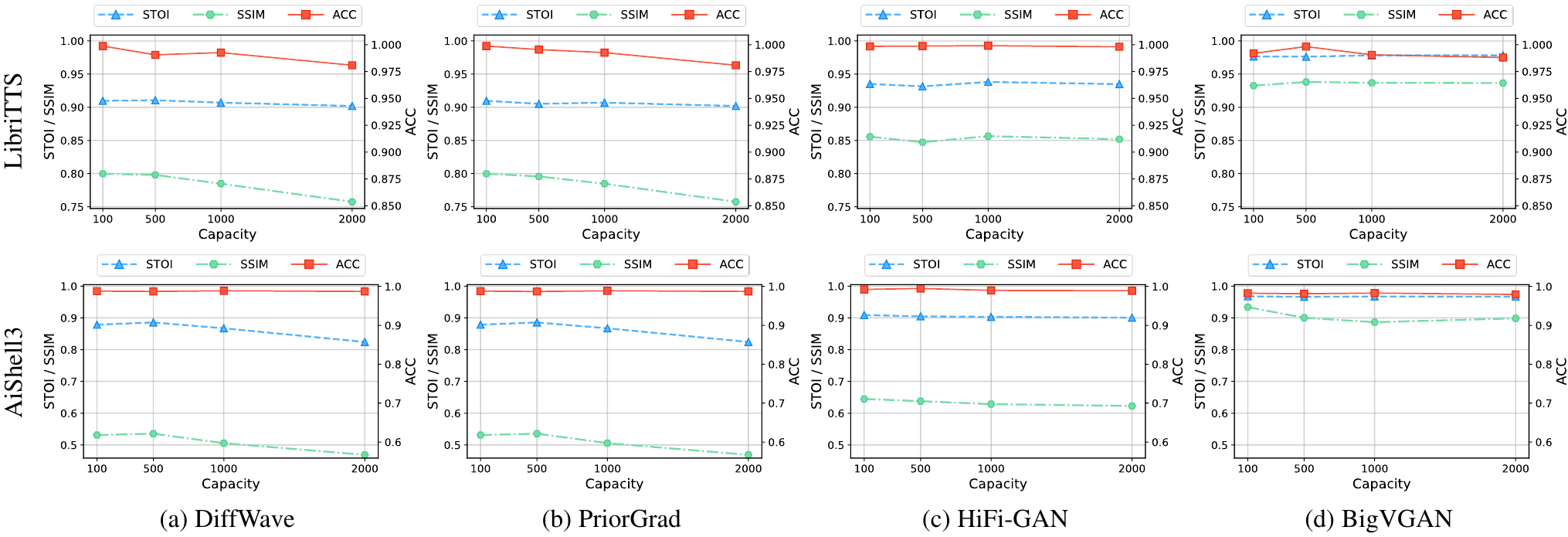}
    \caption{Scalability Analysis Under Varying Watermark Capacities on Out-of-Distribution LibriTTS and AiShell3 Dataset.}
    \label{fig_app_gen_cap}
\end{figure}

\subsection{Generalization Analysis}
\label{sec_app_genera}
To ensure VocBulwark aligns with practical application scenarios and satisfies real-world deployment demands, serving as a complement to Sec.~\ref{sec_eff}.
The experiments assess the method's performance across three critical dimensions: fidelity, capacity, and robustness against diverse distortions.

\textbf{Fidelity.}
Table~\ref{tab_app_fidelity} presents the fidelity performance on LibriTTS and AiShell3 at a capacity of 100 bps. 
In multi-speaker scenarios, VocBulwark maintains high fidelity scores comparable to the clean baselines in LibriTTS. 
Specifically, our DW achieves a PESQ of 2.5057 and STOI of 0.9097, closely tracking the original DiffWave (PESQ of 2.6435, STOI of 0.9232). 
Moreover, in cross-lingual settings, VocBulwark retains the ability to synthesize high-quality watermarked speech in AiShell3. 
Notably, our BVG achieves a PESQ of 3.4325, only marginally lower than the clean baseline of 3.4614. 
This further validates that our acoustic feature embedding successfully guarantees high-quality speech generation.

\textbf{Capacity.}
We also conducted capacity experiments on LibriTTS and AiShell3 with capacities ranging from 100 bps to 2000 bps, as shown in Fig.~\ref{fig_app_gen_cap}. 
The results indicate a solid tolerance for high watermark capacities. 
Specifically, VocBulwark exhibits remarkable scalability on LibriTTS, where it maintains an accuracy exceeding 97\% across different backbones even at a high capacity of 1000 bps while preserving respectable speech quality. 
For AiShell3, VocBulwark effectively adapts to cross-lingual data since accuracy does not significantly drop as capacity increases, and thus ensures high-quality speech generation. 
When the capacity reaches 2000 bps, we observe varying degrees of perceptual quality loss, although the accuracy remains stable without a distinct downward trend. 
Therefore, we consider 2000 bps as the upper bound for maintaining optimal performance, as described in Sec.~\ref{sec_fi_ca}.

\textbf{Robustness Against Standard Attacks.}
We evaluated the robustness of VocBulwark against standard attacks, as presented in Table~\ref{tab_app_std_atk}. The results demonstrate that our method exhibits exceptional resilience across diverse datasets, maintaining superior performance even under high-capacity conditions.
Facing common attacks, VocBulwark does not exhibit vulnerability when applied to OOD datasets.
Instead, it consistently maintains an accuracy exceeding 90\% across the majority of attack scenarios on both LibriTTS and AiShell3.
Furthermore, this robustness is well-preserved even at high capacities, exhibiting only marginal performance degradation. 
Regarding variable-length attacks, VocBulwark demonstrates superior resilience, achieving over 90\% accuracy across diverse backbones. Crucially, even at a high capacity of 2000 bps, the system ensures reliable watermark recovery from fragmented speech, effectively validating our broadcasting concept where local segments retain sufficient information for extraction.

\textbf{Robustness Against Codec Attacks.}
We further evaluated the generalization capability of VocBulwark on OOD datasets, as summarized in Table~\ref{tab_app_Codec_atk}. 
In the realm of traditional Codecs, VocBulwark exhibits exceptional resistance to compression, where it achieves accurate watermark recovery under MP3 and Opus attacks at 64 kbps across both LibriTTS and AiShell3 datasets. 
This robustness persists as watermark capacity increases and thus validates the stability of our method against standard lossy compression. 
Regarding neural Codec regeneration, the strategic advantage of our acoustic feature alignment ensures robust performance on OOD data and enables an average accuracy of approximately 90\% on general-purpose Codecs. 
Our method also effectively withstands regeneration from semantic distillation and disentanglement-based Codecs which inherently impose greater destruction on watermark features. 
Even when capacity escalates to 2000 bps, our approach maintains a solid defense as the GAN-based variant sustains an average accuracy exceeding 80\%.

\textbf{Robustness Against Compound Attacks.}
To rigorously assess stability under complex distortions, we conducted experiments on OOD datasets using compound attacks at high capacities, as detailed in Table~\ref{tab_app_cmp_atk}.
The results demonstrate that our method consistently retains high accuracy when subjected to the superposition of two common attacks, where no significant degradation is observed even as capacity increases.
Our VocBulwark also effectively withstands combinations of common distortions and Codec compression and thus ensures reliable watermark extraction.
Facing the most severe scenario involving dual-Codec compression, which inherently imposes greater destruction on watermark features, VocBulwark demonstrates remarkable durability.
Even at the extreme capacity of 2000 bps, it secures a recovery accuracy exceeding 80\% and thereby validates its viability in rigorous real-world environments.

\textbf{Summary.}
The generalization experiments demonstrate that VocBulwark is highly adaptable to real-world scenarios, satisfying the critical demand for regulating diverse generative models and prolific generated content. 
This capability stems from our proposed Acoustic Feature Alignment and Frame-level Temporal Broadcasting mechanisms, which facilitate the seamless fusion of watermark signals with intrinsic acoustic features during the generation process. 
Consequently, TA endows the host vocoder with a unique identifier, establishing a more reliable and trustworthy mechanism for provenance tracking and content regulation.
Furthermore, to better balance the trade-off among fidelity, robustness, and capacity at the ultra-large capacity of 1000 and 2000 bps, we strategically biased the training process towards watermark recovery.
This approach intentionally prioritizes robustness by sacrificing a degree of fidelity, which explains the observed decline in speech quality metrics while simultaneously yielding superior robustness performance compared to the 500 bps.
Collectively, these design choices empower VocBulwark to deliver high-fidelity watermarked speech generation and sustain robust resilience against diverse attacks, regardless of the complexities in multi-speaker and cross-lingual scenarios.

\begin{table}[t]
\centering
\caption{Quantitative Ablation Study of the Cage Components on LJSpeech.}
\resizebox{0.95\textwidth}{!}{
\begin{tabular}{cccccccccc}
\toprule
Mehod      & STOI~$\uparrow$   & PESQ~$\uparrow$   & SSIM~$\uparrow$   & MCD~$\downarrow$    & DNSMOS~$\uparrow$ & ACC~$\uparrow$ & $\Delta \text{ACC}$  & Params.~$\downarrow$ & FLOP~$\downarrow$  \\ 
\midrule
VocBulwark[DW] & 0.9605 & 3.3398 & 0.8519 & 6.3682 & 3.2727 & 0.9998 & - & 1.66 M  &  3.59$\times 10^{11}$   \\
\multicolumn{1}{l}{\hspace{2em} w/o TA}  & 0.9582 & 3.4104 & 0.8422 & 6.2832 & 3.2648 & 0.7230 & -0.2768 & 869.04 M  &  3.50$\times 10^{11}$  \\
\multicolumn{1}{l}{\hspace{2em} w/o Alignment}  & 0.9578 & 3.2879 & 0.8356 & 6.5269 & 3.1690 & 0.9884 & -0.0114 & 58.57 M  &  3.59$\times 10^{11}$  \\
\multicolumn{1}{l}{\hspace{2em} w/o Broadcasting}  & 0.9588 & 3.1434 & 0.8516 & 6.2740 & 3.2660 & 0.9992 & -0.0006 & 30.36 M  &  3.59$\times 10^{11}$  \\
\multicolumn{1}{l}{\hspace{2em} w/o GSC}    & 0.9666 & 3.3768 & 0.8658 & 6.5514 & 3.3108 & 0.6902 & -0.3096 & 1.85 M  &  3.59$\times 10^{11}$  \\ 
\multicolumn{1}{l}{\hspace{2em} w/o Gated}  & 0.9606 & 3.2760 & 0.8552 & 6.4164 & 3.2530 & 0.9898 & -0.0100 & 1.40 M  &  3.58$\times 10^{11}$  \\
\multicolumn{1}{l}{\hspace{2em} w/o AGOC}  & 0.9690 & 3.4120 & 0.8696 & 6.5728 & 3.3196 & 0.6844 & -0.3154 & 1.66 M  &  3.59$\times 10^{11}$  \\
\bottomrule
\end{tabular}
}
\label{tab_app_ablation}
\end{table}

\section{Ablation Study}
As shown in Table~\ref{tab_app_ablation}, we perform a comprehensive ablation study to validate the contribution of key components (TA and Cage architecture) and the training strategy (AGOC). 

\textbf{Effectiveness of TA Components.} 
To rigorously validate the efficacy of the TA, we conducted a comprehensive ablation study targeting its core mechanisms.
We first replaced the TA with Groot's~\cite{liu2024groot} embedding strategy of Input Modification (w/o TA). 
Specifically, we utilized Groot's encoder to embed the watermark directly into the input of the diffusion model, while training the watermarking via the VocBulwark strategy.
As evidenced in Table~\ref{tab_app_ablation}, this noise-space embedding fails to effectively fuse the watermark with the generative model, causing a sharp decline in extraction accuracy.
Moreover, compared to our lightweight Additional-Parameter Injection, this approach incurs an exorbitant computational cost, requiring 869.04 M parameters.
Moreover, we validated the Acoustic Feature Alignment by replacing it with a naive mapping strategy that projects watermark features directly to match the temporal dimension (w/o Alignment).
The results indicate that forcing temporal alignment without respecting acoustic attributes degrades generation fidelity and demands a significantly higher parameter overhead (58.57 M).
Finally, we ablated the Frame-level Temporal Broadcasting (w/o Broadcasting) by utilizing FC layers after the PFP to match the hidden feature dimensions.
While this configuration compromises fidelity, its most critical failure lies in robustness against desynchronization. Lacking the time-invariant properties conferred by broadcasting, the model becomes extremely vulnerable to variable-length manipulations, with accuracy plummeting to 68.68\% under Time Stretching and 69.32\% under Cropping.
In a nutshell, these results further validate that our TA achieves a trade-off between fidelity and robustness while utilizing significantly fewer parameters, highlighting that acoustic feature alignment enhances generative capabilities and the broadcasting mechanism confers essential resilience against variable-length attacks.

\textbf{Effectiveness of Cage Components.} 
Regarding GSCM of Cage, replacing the GSC with a standard Conv1d resulted in a significant decline in performance, with accuracy dropping by 30.96\% to 69.02\%. 
This empirical evidence underscores the critical role of the GSC in effectively disentangling watermark features from complex acoustic backgrounds, thereby ensuring precise and robust recovery.
In addition, we ablated the gating mechanism within the GSCM, utilizing only DSC as the backbone. 
While removing the gating mechanism yields marginal gains in STOI and SSIM, it simultaneously degrades critical perceptual metrics (PESQ, DNSMOS, MCD) and reduces extraction accuracy by 1.00\%.
Consequently, we retain the gating mechanism as it effectively harmonizes the trade-off, ensuring optimal auditory naturalness and near-perfect watermark recovery.

\textbf{Effectiveness of Accuracy-Guided Optimization Curriculum.} 
The ablation of the Accuracy-Guided Optimization Curriculum (w/o AGOC) yields the most significant performance degradation in terms of watermark extraction, with the accuracy plummeting by 31.54\% to 0.6844.
Interestingly, this configuration achieves the highest scores in perceptual metrics (PESQ of 3.4120 and STOI of 0.9690).
This phenomenon empirically validates the design rationale behind AGOC, which aims to mitigate the gradient dominance of perceptual reconstruction in the early training phase.
By prioritizing the establishment of watermark injectability over immediate generation fidelity, AGOC utilizes accuracy-guided dynamic re-weighting to ensure that conflicting objectives are effectively harmonized.

\section{Visualization of Fidelity Compared With Baselines}
\begin{figure}[t]
    \centering
    \includegraphics[width=0.95\linewidth]{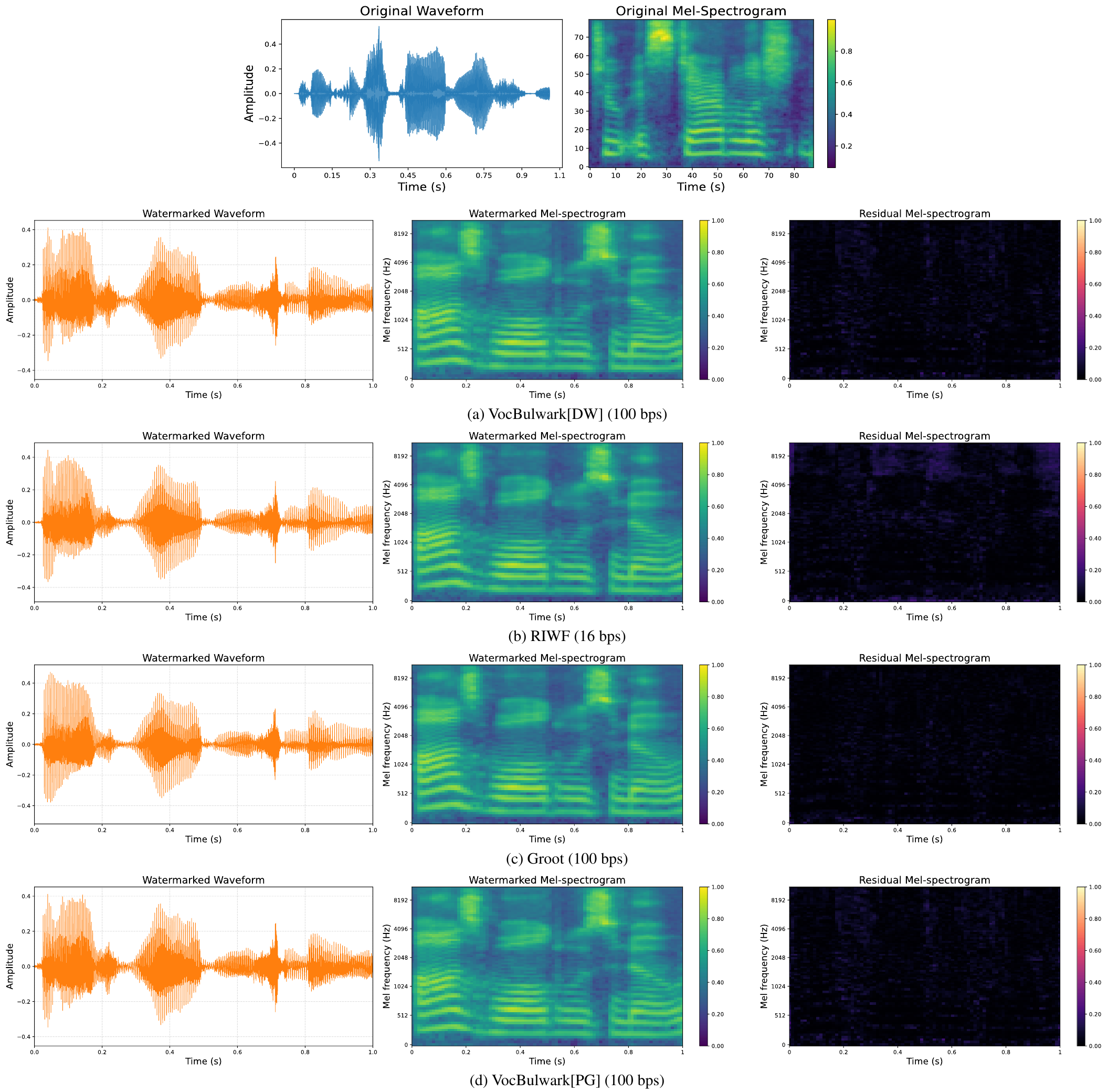}
    \caption{Qualitative visualization of Fidelity on Diffusion Models. The top row displays the waveform and Mel-spectrogram of the original speech. The subsequent rows present the waveforms, Mel-spectrograms, and residual maps of watermarked speech generated by various watermarking methods on Diffusion Models.}
    \label{fig_app_vis_dm}
\end{figure}

\begin{figure}
    \centering
    \includegraphics[width=0.95 \linewidth]{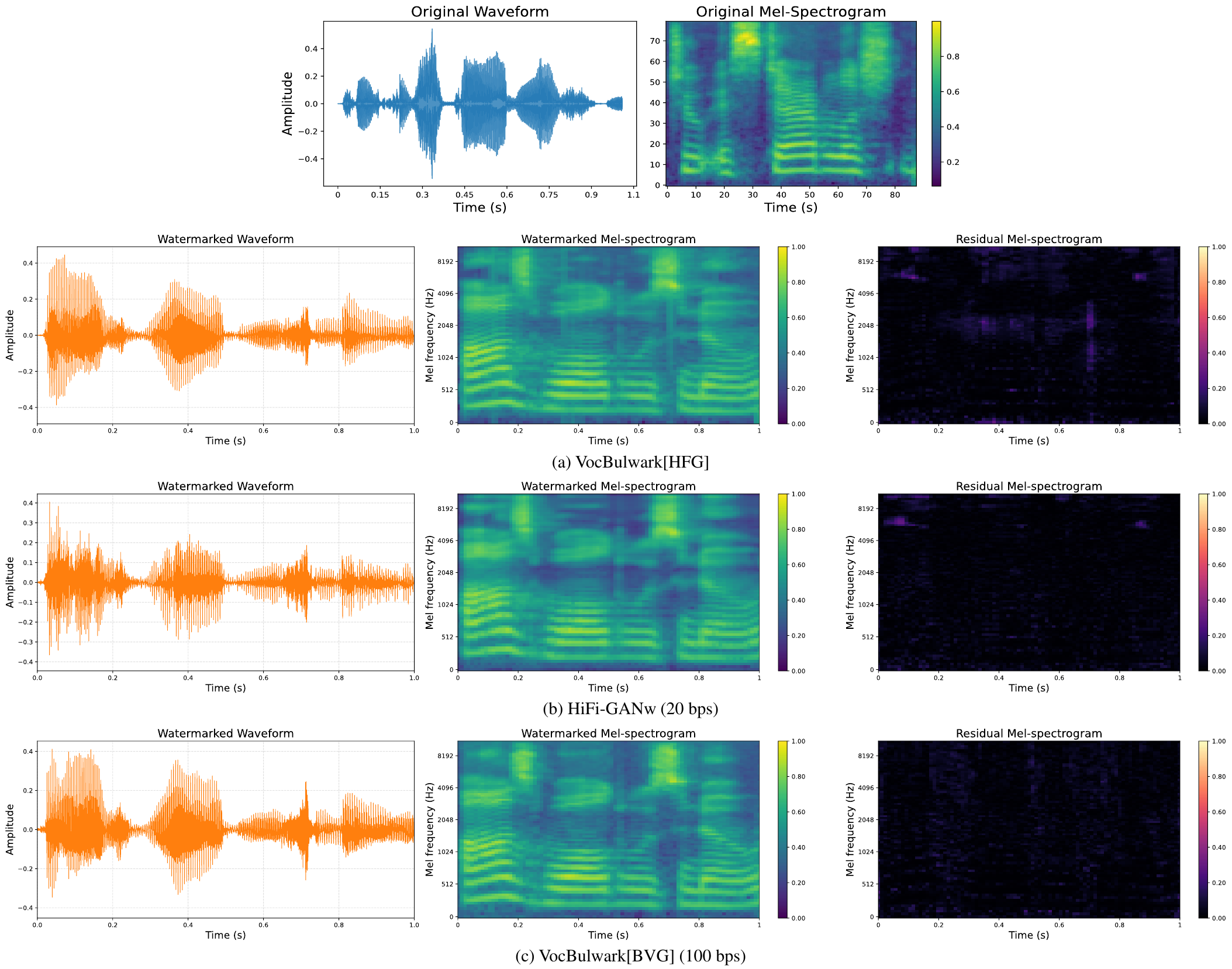}
    \caption{Qualitative visualization of Fidelity on GANs. The top row displays the waveform and Mel-spectrogram of the original speech. The subsequent rows present the waveforms, Mel-spectrograms, and residual maps of watermarked speech generated by various watermarking methods on GANs.}
    \label{fig_app_vis_gan}
\end{figure}

To further demonstrate the fidelity of our watermarking approach, we visualize the waveforms, Mel-spectrograms, and spectral residual maps of watermarked speech generated by VocBulwark and the baselines. 
Specifically, Fig.~\ref{fig_app_vis_dm} and Fig.~\ref{fig_app_vis_gan} present the detailed qualitative results on Diffusion models and GAN-based architectures, respectively.
To ensure a fair comparison, the speech samples used for visualization were randomly sampled from the test set on LJSpeech.

As evidenced in Fig.~\ref{fig_app_vis_dm}, VocBulwark introduces minimal degradation to the original speech. 
In the time domain, our DW exhibits excellent envelope consistency with significantly reduced amplitude fluctuations. 
Notably, around 0.05s, our method effectively suppresses anomalous peaks, maintaining a closer alignment with the ground truth. 
Furthermore, the residual spectrograms reveal that our watermark exerts negligible distortion in the frequency domain. 
In stark contrast to the distinct block-like artifacts observed in RIWF, our approach achieves exceptional perceptual quality.
Fig.~\ref{fig_app_vis_gan} corroborates the generalization capability of VocBulwark across GAN-based neural vocoders. Consistent with the observations in diffusion models, the watermarked speech generated by our method maintains excellent alignment with the original waveform envelopes and spectral structures.

\section{Discussion}
While images benefit from spatial redundancy, speech represents a one-dimensional temporal signal that exhibits higher sensitivity and significantly lower redundancy.
Such intrinsic properties inevitably exacerbate the difficulty of embedding imperceptible and robust watermarks.
Additionally, most existing vocoders synthesize waveforms from Mel-spectrograms by recovering phase information rather than operating within the latent spaces commonly utilized in image generation.
This distinct generation paradigm inherently complicates the watermark embedding process and necessitates a more rigorous trade-off among fidelity, capacity, and robustness.
To address these domain-specific challenges, VocBulwark employs an additional-parameter injection strategy. 
By effectively integrating the watermark with intrinsic acoustic attributes, our approach achieves a balanced trade-off and ensures the watermark remains imperceptible yet robust within the generated speech.

Furthermore, we observe that the underlying architecture of the generative model fundamentally influences watermarking performance.
Specifically, diffusion models foster greater robustness through their iterative denoising training processes, while GAN-based architectures demonstrate superior efficacy under high capacity.
This distinction suggests that future work could explore hybrid frameworks that leverage the complementary strengths of both architectures to optimize performance across different vocoders.
Moreover, the rapid evolution of neural Codecs poses emerging threats to watermarking security.
Modern Codecs are advancing towards lower bitrates and sophisticated disentanglement, where the focus shifts from simple quantization to the complex separation of semantic and acoustic features.
This progression aggressively strips away non-essential information and thus threatens the survival of hidden watermarks.
Although VocBulwark mitigates this issue to a certain extent, ensuring watermark survivability against increasingly advanced compression and disentanglement techniques remains a pivotal direction for future investigation.

%%%%%%%%%%%%%%%%%%%%%%%%%%%%%%%%%%%%%%%%%%%%%%%%%%%%%%%%%%%%%%%%%%%%%%%%%%%%%%%
%%%%%%%%%%%%%%%%%%%%%%%%%%%%%%%%%%%%%%%%%%%%%%%%%%%%%%%%%%%%%%%%%%%%%%%%%%%%%%%

\end{document}